\documentclass[twoside,11pt]{article}

\usepackage{amsmath}
\usepackage{amssymb}
\usepackage{amsxtra}

\usepackage[dvips]{graphics}

\usepackage{hyperref}

\hypersetup{
    colorlinks = true,
    linkcolor = green,
    anchorcolor = red,
    citecolor = blue,
    filecolor = red,
    pagecolor = red,
    urlcolor = red
}

\hypersetup{
    pdftitle =
        Topological strings in generalized complex space,
    pdfauthor =
        Vasily Pestun,
    pdfsubject=
        hep-th,
    pdfkeywords =
        generalized complex structure
}

\newcommand {\la} {\left \langle}
\newcommand {\ra} {\right \rangle}

\newcommand {\CalA} {\mathcal A}
\newcommand {\CalD} {\mathcal D}

\newcommand {\CalF} {\mathcal F}

\newcommand {\CalI} {\mathcal I}
\newcommand {\CalJ} {\mathcal J}
\newcommand {\CalO} {\mathcal O}

\newcommand {\CalN} {\mathcal N}

\newcommand {\CalL} {\mathcal L}

\newcommand {\CalM} {\mathcal M}

\newcommand {\BR}   {\mathbb R}
\newcommand {\BZ}   {\mathbb Z}
\newcommand {\BC}   {\mathbb C}

\newcommand {\al} {\alpha}

\newcommand {\be} {\beta}

\newcommand {\de} {\delta}

\renewcommand{\Im} {\mathrm Im}

\newcommand {\p} {\partial}

\DeclareMathOperator{\End}{End}

\DeclareMathOperator{\tr} {tr}

\DeclareMathOperator{\HH} {HH}
\DeclareMathOperator{\I} {Im}
\DeclareMathOperator{\R} {Re}
\DeclareMathOperator{\Maps} {Maps}

\newcommand{\Spin}{\mathop{\rm Spin}\nolimits}

\numberwithin{equation}{section}

\begin{document}

\title{
\begin{flushright}
\vspace{-20mm}
\small PUTP-2188 \\
ITEP-TH-104/05
\vspace{10mm}
\end{flushright}
{\huge{\bf{Topological strings\\in generalized complex space}}}
  \author{\sc{\large{ Vasily Pestun}\footnote{On leave of absence from ITEP, Moscow, 117259, Russia }}\\
    {\it Physics Department, Princeton University, Princeton NJ 08544, USA}\\
    {\small{e-mail:}} {\tt {\small{pestun@princeton.edu}}}}
  \date{March 2006} \vskip5mm}

\maketitle

\thispagestyle{empty}

\begin{abstract}
A two-dimensional topological sigma-model
on a generalized Calabi-Yau target space $X$ is defined.
The  model is constructed in Batalin-Vilkovisky
formalism using only a generalized complex structure $\CalJ$ and a pure spinor $\rho$ on
$X$. In the present construction the
algebra of $Q$-transformations automatically closes off-shell,
the model transparently depends only on $\CalJ$, the
algebra of observables and correlation functions for
topologically trivial maps in genus zero are
easily defined.
The extended moduli space appears naturally. The familiar action of the twisted $\CalN=2$ CFT can
be recovered after a gauge fixing. In the open case, we consider an
example of generalized deformation of complex structure by
a holomorphic Poisson bivector $\beta$ and recover holomorphic
noncommutative Kontsevich $*$-product.
\end{abstract}

\section{Introduction}

The topological~\cite{Witten:1988xj,Witten:1988ze} A/B-sigma-models~\cite{Witten:1991zz} were introduced by Witten
as twists of the $\CalN=2$ two-dimensional supersymmetric conformal field theory
of maps from the worldsheet $\Sigma$ into the target space $X$.
Topological strings~\cite{Bershadsky:1993cx} are obtained by
coupling topological sigma-models to  topological 2D gravity on $\Sigma$, so that in the path integral one integrates over the
moduli space of complex structures on~$\Sigma$.
See~\cite{MR2003030} for references and an extensive review of the
progress in the subject.

The topological A-model can be defined for any almost Kahler
manifold $X$. An almost Kahler structure is a pair $(\omega,J)$ of a symplectic
structure $\omega$ and a compatible almost complex structure $J$.
In the Batalin-Vilkovisky approach, the almost complex structure $J$ is used for the gauge fixing of
the A-model, so the correlation functions depend only on the
symplectic structure $\omega$. The observables $t$ of the A-model
are identified with the de Rham cohomology classes $t \in
H^{*}(X,\BC)$. The A-model free energy $\CalF_{g}(t)$ is the
generating function of genus $g$ Gromov-Witten invariants of
$X$~\cite{Witten:1990hr,MR1363062}.

On a Kahler manifold, the space of physical observables
$\phi(x)_{i_1\dots i_p\bar i_1\dots \bar i_q}\chi^{i_1}\dots
\chi^{i_p} \chi^{\bar i_1} \dots \chi^{\bar i_q}$ of the
$A$-model~\cite{Witten:1991zz} is conveniently graded by a pair
$(p,q)$ according to the Hodge decomposition of the de Rham
cohomology $H^{k}(X,\BC) = \oplus_{p+q = k} H^{p,q}(X,\BC)$.
The cohomology class of symplectic structure $\omega$ can be
deformed by an element of  $H^{2}(X,\BC)$, which corresponds to an
observable of degree $p+q = 2$. Physically, such observables give
rise to ghost number zero deformations  of the
action~\cite{Witten:1991zz} for the $A$-model. The deformations
corresponding to the observables of degree $p+q=2$ will be called
\emph{geometrical deformations.} The hermitian observables of type
$(1,1)$ are deformations of the Kahler structure. If we include
the B-field then the hermitian condition can be dropped. In
other words, the real $B$-field and the real symplectic structure
$\omega$ can be combined together into a complex two-form  $\omega +
iB$.

If $H^{2,0}(X,\BC)$ is trivial, then deformations of type $(1,1)$
are the only geometrical deformations of the $A$-model. However, if $H^{2,0}(X,\BC)$
is not trivial, then there are also observables of type $(2,0)$ and
$(0,2)$.  Let us call the moduli space generated by the observables of degree
(1,1) --- \emph{the ordinary geometric} moduli space,
for the degrees $(2,0), (1,1), (0,2)$ --- \emph{the
geometric} moduli space, and for all $(p,q)$   ---
\emph{the extended} moduli space following~\cite{Witten:1991zz}.

Now consider the topological B-model on a Calabi-Yau manifold $X$.
The B-model couples to a complex structure~$J$ on~$X$. Its
observables~$\phi_{\bar j_1\dots \bar j_q}^{i_1\dots i_p}(x) \theta^{i_1}\dots
\theta^{i_p} \eta^{\bar j_1} \dots \eta^{\bar j_q}$~\cite{Witten:1991zz}
of type $(-p,q)$ are identified with the
Dolbeault cohomology classes $H^q(\Lambda^p
TX^{1,0})$.\footnote{For convenience we will call sections of
$\Lambda^p (TX^{1,0})\otimes \Lambda^q (T^*X^{0,1})$ by $(-p,q)$
forms.} The ordinary geometric moduli space of complex structures
is generated by Beltrami differentials $\mu^{i}_{\bar j}$.
They deform the Dolbeault differential  $\bar \p_{\bar j} \to \bar \p_{\bar j} + \mu_{\bar j}^{i} \p_i$.
The genus $g$ free energy $\CalF_{g}$ of the B-model does not
have such a clear geometrical description as in the case of the $A$-model.
For the recent mathematical progress in definition of the B-model
see~\cite{Costello:2004ei,Costello:2005cx} and~\cite{Losev:0506039,Gerasimov:2004yx}.
In genus zero the extended moduli space of the B-model
was studied in~\cite{MR1609624,MR1866443,MR1911737}.

Again, the observables of type $(-1,1)$ correspond to ordinary
geometrical deformations of complex structure generated
by Beltrami differentials $\mu_{\bar j}^i$.
If $H^0(\Lambda^{2} TX^{1,0})$ or $H^{0,2}(X)$ are nontrivial,
then there exist \emph{generalized} deformations of complex structure
generated by a holomorphic Poisson bivector $\beta^{ij}$ or by a closed
two-form $B_{\bar i\bar j}$.

\begin{align}
\label{ord_gen_wit}
\boxed{\stackrel{\text{\emph{ordinary} geometric moduli}}{(p,q)=(1,1)}} \subset \boxed{
\stackrel{\text{(Hitchin's  \emph{generalized}) geometric moduli}}{p+q=2}}
\subset \boxed{\stackrel{\text{\emph{extended} moduli}}{\text{all } (p,q)}}
\end{align}

 The geometrical meaning of these
deformations was unclear in~\cite{Witten:1991zz,Bershadsky:1993cx}.
The missing notions appeared in the studies of the homological mirror
symmetry~\cite{MR1403918}
and the generalized complex geometry~\cite{Hitchin:2}.\footnote{Let us remark that these
deformations were often neglected in the literature, since usually
the target space $X$ was taken to have $b_1(X)=0$.}

In the framework of the Kontsevich homological mirror symmetry
conjecture~\cite{MR1403918},
we need to consider not only closed strings, but
also open strings together with all possible branes.
Physically, branes are the boundary conditions for the open
strings. Mathematically, branes are described by certain derived categories.
For example, in the case of the B-model,
branes could be holomorphic vector bundles supported on holomorphic
submanifolds. The full category of B-branes is the bounded derived
category $\CalD^b(X)$ of coherent sheaves on $X$.
Physically, the worldsheet path integral defines the structure of  algebra on the
space of open string states $\CalA$. For example, the product $ \CalA \otimes \CalA \to \CalA$
can be defined in terms of the open string three-point
function. The worldsheet action in the path integral can be
deformed by closed string observables. By definition, this deforms
the open string algebra. Therefore, deformations of the open
string algebra (modulo some symmetry) are described by  the closed
string
states~\cite{Lazaroiu:2003md,Aspinwall:2004jr,Sharpe:2003dr,Hofman:2000ce,
Hofman:2002cw,Costello:2004ei,Costello:2005cx}.

Mathematically, these deformations
of the category of branes are computed by the Hochschild
cohomology~$\HH(\CalD^b(X))$~\cite{MR2062626,MR1855264,Kapustin:2004df,Hofman:2002cw,Hofman:2000ce,Kajiura:2005sn,MR1911737}. In~\cite{Kapustin:2004df} it was argued,
that computation of $\HH(\CalD^b(X))$ can be reduced to the computation
of the Hochschild cohomology of the single space-filling brane.
The algebra of observables on such a brane is just an
algebra of holomorphic functions~$\CalO(X)$. Thus, the hard
computation of~$\HH(\CalD^b(X))$ reduces to an easier computation
of $\HH(\CalO(X))$ and gives the result  \cite{Kapustin:2004df,MR1390671}
\[\HH(\CalO(X)) = \oplus
H^q(\Lambda^p(TX^{1,0})).\]
This is precisely the moduli space of the closed B-model~\cite{Witten:1991zz}.
Given this observation, the geometrical nature of
$(-2,0)$ and $(0,2)$ deformations of the closed B-model might be
interpreted in terms of equivalent deformation of the open
B-model algebra. It was claimed
in~\cite{Caldararu:2003kt,Kapustin:1999di} that $(0,2)$
deformations by a closed two-form $b_{\bar i \bar j}$ transform a
sheaf structure on a brane into a more general gerby structure. On
the other hand, $(-2,0)$ deformations by a holomorphic
bivector~$\beta^{ij}$ deform the ordinary product of functions on
a brane into the noncommutative Kontsevich
$*$-product~\cite{MR2062626,MR1855264,Hofman:2002cw,Losev:1997tp,Kapustin:2003sg}.
See~\cite{Douglas:2001ba} for a review of noncommutative field theories,
and~\cite{ben-bassat-2005,ben-bassat-2006-56,block-2005} for study of mirror symmetry in the
context of generalized complex structures.

The notion of a generalized complex structure and generalized Calabi-Yau manifold
was defined by Hitchin~\cite{Hitchin:2,Hitchin}
and then fully developed in Gualtieri's thesis~\cite{GCS}.
There are two ideas behind the notion of generalized complex structure. The first, coming
from theory of constrained systems,
is to generalize structures on the tangent bundle $TX$ to structures on the direct sum of the tangent
and the cotangent bundle $E = TX \oplus T^*X$ and consider Dirac
structure on $E$~\cite{MR998124}. The second, coming from string
theory, is to extend the diffeomorphism group by action of B-field.

An \emph{almost complex structure} is a section $I$ of $\End(TX)$ such that
$I^2=-1$. Similarly, an \emph{almost generalized complex structure} is a
section $\CalJ$ of $\End(TX\oplus T^*X)$ such that $\CalJ^* = -\CalJ, \CalJ^2=-1$.
An almost complex structure is integrable if $+i$-eigenbundle of $J$
(holomorphic subbundle of $TX$) is involutive under the Lie
bracket on $TX$. Similarly, an almost generalized complex structure is
integrable if $+i$-eigenbundle $L$ of $\CalJ$ (holomorphic
subbundle of $TX\oplus T^*X$) is involutive under the Courant
bracket~\cite{MR998124}.

The generalized complex geometry incorporates symplectic
structures and ordinary complex structures as particular cases.
Therefore, the topological A-model and the topological B-model
could be particular cases of a certain generalized topological
model~\cite{Kapustin:2003sg,Kapustin:2004gv,Kapustin:2005uy}.
This topological sigma-model of maps from $\Sigma$ to $X$
depends on a generalized complex structure $\CalJ$, so let us call it \emph{the topological $\CalJ$-model}.
See~\cite{Kotov:2004wz} for the analogue of the present
construction in the real case and~\cite{Alekseev:2004np}
for studies of the current algebra associated with $\CalJ$.

If $\CalJ$ is an ordinary symplectic(complex) structure, let us call it
to be of A(B)-type. Gualtieri~\cite{GCS} shows that deformations of a generalized
complex structure at the symplectic point (A-type) are
parameterized by $H^2(X,\BC)$, while deformations in the complex point (B-type)
are parameterized by $\oplus_{p+q=2} H^{q}(\Lambda^pTX^{1,0})$.
This is the space of $p+q=2$ deformations
of the topological A/B-model~\cite{Witten:1991zz}.
Thus it is very natural to suggest that the topological A/B-model is a particular
case of the generalized $\CalJ$-model.

Then one can also ask the following question.
If the Hitchin's \emph{generalized} complex geometry
corresponds to  $p+q=2$ deformations of the A/B-model, then what
could be the \emph{extended} complex geometry that corresponds to arbitrary $(p,q)$
deformations?
nn
The answer is known under the name BV geometry. It first appeared in papers of Batalin and
Vilkovisky~\cite{Batalin:1981jr,Batalin:1984jr}, who suggested a powerful
generalization of BRST quantization method for the case when gauge symmetries are
reducible. A clear geometric interpretation of the BV formalism
was given in~\cite{Witten:1990wb,Schwarz:1992nx,Schwarz:2000ct}.
A general method to construct a topological sigma model for a
BV target manifold was suggested in~\cite{MR1432574} by Alexandrov, Kontsevich, Schwarz and Zaboronsky (AKSZ).
They illustrated the method by example of Chern-Simons theory and the topological
A(B)-model. Later AKSZ method was used by Kontsevich~\cite{MR2062626}
to find the $*$-product formula in the context of deformation quantization,
see also~\cite{MR1779159,MR1854134,MR2104442}.

This is a summary of the steps to get the $\CalJ$-model
\[
\boxed{\text{generalized CY structure}} \stackrel{1}{\longrightarrow} \boxed{\text{Lie bialgebroid}}
\stackrel{2}{\longrightarrow} \boxed{\text{BV geometry}}
\stackrel{3}{\longrightarrow}
\boxed{\text{$\CalJ$-model}}.
\]

1. A generalized complex structure $\CalJ$ defines decomposition of
the bundle $E = (TX \oplus T^*X)\otimes C$ into the
$+i$-eigenbundle $L$ and its conjugate $L^*~\simeq \bar L$, so $E = L \oplus
\bar L$~\cite{Hitchin:2,GCS}. Each of the bundles $L$ and $\bar L$
has a structure of Lie algebroid~\cite{MR0216409,MR998124,Roytenberg:thesis,MR1472888}.
The Lie bracket $\{, \}: \Gamma(L) \otimes \Gamma(L) \to \Gamma(L)$ is  equal to the
restriction of the Courant bracket from  $TX \oplus T^*X$ to
$L$. The  Lie bracket on $L^*$ is also the restriction of the Courant bracket.
 The Lie algebroid differential
$\bar \p: \Gamma(\Lambda^k L^*) \to \Gamma(\Lambda^{k+1} L^*)$ is
canonically defined by the Lie bracket on $L$. It satisfies $\bar
\p^2 = 0$.
The pair $(L,L^*)$ has the structure of Lie bialgebroid $(L,L^*)$~\cite{MR1262213},
which means that the operator $\bar \p$, defined by the Lie bracket on $L$,
satisfies the Leibnitz's rule for the Lie bracket on $L^*$.

2. A BV manifold $(M, \omega, \rho)$
is generally defined as a supermanifold $M$ equipped with a nondegenerate
symplectic structure $\omega$ and a measure $\rho$ such that the
corresponding odd Laplacian $\Delta$ squares to zero $\Delta^2 = 0$~\cite{Schwarz:1992nx,Schwarz:2000ct}.
Here the Laplacian of a function $f$ is the divergence $\Delta f := div X_f$
of the vector field $X_{f}$ generated by $f$,
where the divergence of a vector field $X$ is defined by the formula $\rho \, div X := d(i_X
\rho)$. The Poisson bracket $\{,\}$ and the solution $S$ of the BV classical master equation
$\{S , S\}=0$ define the operator $Q$ on functions on $M$ as
$Q f := \{S,f\}$. The operator $Q$ is a derivation of the Poisson bracket $\{,\}$
and satisfies $Q^2 = 0$. The space of functions $\CalA :=C^{\infty}(M)$ on such a manifold $M$
is a differential BV algebra~\cite{MR1919435,MR1675117,MR1854134,Getzler:1994yd}.
This algebra has an odd
Lie bracket $\{,\}: \CalA \otimes \CalA \to \CalA$,
canonically defined by the odd Poisson structure $\{f,g\}:=\p_i f \omega^{ij} \p_j
g$, and the differential $Q: \CalA \to \CalA$, defined by the solution
of the classical master equation $Qf : = \{S,f\}$. The
Poisson bracket in $\CalA$ is generated by the BV Laplacian
$\Delta$ as  $\{a,b\} = (-1)^{|a|}( \Delta(ab) - (\Delta a)
b - (-1)^{a} a (\Delta b))$.

Now consider the supermanifold $N = \Pi L$, which is the total space of
the bundle $L$ with parity reversed on the fibers.
In other words, coordinates in the fibers are odd (fermionic) variables.
The space of functions $A=C^{\infty}(N)$ is naturally identified
with the space of sections of $\oplus_{k} \Lambda^{k}L^*$.
The Lie algebroid differential $\bar \p: \Gamma(\Lambda^k L^*) \to \Gamma(\Lambda^{k+1}L^*)$ is
equivalent to the BV differential $Q$ of degree 1 on $\CalA$.
The Lie algebroid bracket $\{ , \}: \Gamma(L^*) \otimes \Gamma(L^*) \to \Gamma(L^*)$
is equivalent to the BV bracket $\CalA \otimes \CalA \to \CalA$ of degree -1.
The generalized CY condition is equivalent to the existence of BV measure $\rho$.
It allows us to define the BV Laplacian $\Delta$. In terms of
the Lie algebroid  one has $\Delta = \p$, where $\p$
is the generalized $\p$-divergence $\Gamma(\Lambda^{k} L^*) \to
\Gamma(\Lambda^{k-1}L^*)$. Actually, the generators of the
Poisson bracket on $\Pi L$ are in one to one correspondence with flat Lie
algebroid $L^*$-connections on the top external power $\Lambda^{\deg L} (L^*)$ --
generalized divergence operators~\cite{MR1625610,MR1906481,MR1675117,MR1362125}.

Thus for any generalized CY manifold there is a corresponding BV
supermanifold $N = \Pi L$ equipped with a homological differential $Q$.
Assume that the odd Poisson structure on $N$ is invertible, so $N$
is a symplectic manifold. If it is not,
then we will take symplectic realization $M$ of $N$~\cite{MR1747916,MR866024,MR854594}.
Since the odd vector field $Q$ preserves the symplectic structure, the form $i_{Q}
\omega$ is closed. If it is also exact, then $i_{Q} \omega = dS$.
The function $S$ is a Hamiltonian function for the vector field
$Q$, so $Q = \{S, \cdot \}$.

3. Given the BV target manifold $N$ and the Hamiltonian function $S$ that satisfies
$\{S,S\}=0$, the topological $\CalJ$-model is obtained along the lines
of~\cite{MR1432574,MR1854134}. The BV fields of the model are maps from $\Pi
T\Sigma$ to $N$, where the supermanifold $\Pi T\Sigma$ is the total
space of the tangent bundle of a Riemann surface $\Sigma$ with
parity reversed on the fibers. Let $X^i$ and $p_i$ be local
coordinates on $N$, which are canonically conjugate with respect to the BV odd
symplectic structure. Then the BV action of the model is
simply
\begin{align}
    S = \int_{\Pi T\Sigma} p_i dX^i + S(p_i,X^i).
\end{align}
The physical action of the sigma-model is obtained after the gauge fixing of this BV
master action. See~\cite{Kotov:2004wz,Ikeda:2002qx,Calvo:2005ww} on the Poisson
sigma-model.

In section~\ref{BValgebra} we consider in more details the
algebraic and geometric structures mentioned above. In section~\ref{2Dtheory}
we will construct the $\CalJ$-model.
In section~\ref{StarProduct} we will compute the correlation functions
on the boundary of the disk for the $\CalJ$-model obtained by a finite deformation of
the B-model by a holomorphic Poisson bivector~$\beta^{ij}$ and reproduce the Kontsevich
$*$-product formula. Section~\ref{Conclusion} concludes the paper.

\section{A differential BV algebra $\CalA$ of a generalized CY manifold \label{BValgebra}}

A \emph{generalized complex structure}~\cite{GCS,Hitchin:2} on a manifold $X$
is a section of $\CalJ \in \End(TX\oplus T^*X)$
such that $\CalJ^*=-\CalJ$
and $\CalJ^2 = -1$ with a certain integrability condition: the $+i$-eigenbundle
$L \subset (TX \oplus T^*X) \otimes \BC$ is involutive
with respect to Courant bracket~\cite{MR998124}
\begin{align}
  [ X +  \xi, Y + \eta ] = [X,Y] + L_{X} \eta - L_Y \xi - \frac 1 2 d( i_X \eta - i_Y \xi)
\end{align}
where  $X,Y \in \Gamma(TX)$ and  $\xi,\eta \in \Gamma(T^*X)$.

Then $(TX \oplus T^*X)\otimes \BC = L \oplus \bar L$,
where $\bar L$ can be identified with $ L^*$, since $L$ is maximal
isotropic\footnote{with respect to the canonical metric on $(TX \oplus T^*X)\otimes \BC$} and $L \cap \bar L =0$.

The bundle $L$ has a structure of a \emph{Lie
algebroid}~\cite{GCS,Hitchin:2,MR1472888,MR1906481,Roytenberg:thesis}.
The \emph{Lie algebroid bracket} is  the restriction
of the \emph{Courant bracket} on $L$. The \emph{anchor map} $a: L \to TX$ is
the restriction of the projection $TX\oplus T^*X \to TX$.

Generally speaking, the structure of Lie algebroid on a bundle
$L$ can be conveniently described by an odd 2-nilpotent vector field $Q$ of degree one
on  $\Pi L$~\cite{MR1432574}. Let $(x^{\mu}, \psi^a)$ be coordinates in $\Pi L$.
Functions $f \in C^{\infty}(\Pi L)$ can be expanded in $\psi^a$
\begin{align}
  \label{eq:f_x_psi}
  f(x,\psi) = \sum \frac 1 {k!} f_{a_1,\dots,a_k}(x) \psi^{a_1}\dots \psi^{a_k}
\end{align}
and the coefficients $f_{a_1,\dots,a_k}$ are identified with sections of $\Lambda^k
L^*$, so $C^{\infty}(\Pi L) = \Gamma(\Lambda^\bullet L^*)$.

The vector field $Q$ defines the operator, which acts on the
space of functions $C^{\infty}(\Pi L) = \Gamma(\Lambda^\bullet L^*)$ as
Lie derivative.
Note that any vector field $Q$ of degree one can represented as
\begin{align}
  \label{eq:Q_coordinates1}
  Q = e^{\mu}_a \psi^a \frac {\partial} {\partial x^{\mu}} - \frac 1 2 f^{a}_{bc} \psi^{b} \psi^{c} \frac {\partial}{\partial
  \psi^a}
\end{align}
it corresponds to the Lie algebroid differential
\begin{equation}
\label{eq:Lie_algebroid_complex}
\cdots \stackrel{\bar \p }{\to}  \Lambda^k L^*
\stackrel{\bar \p}\to \Lambda^{k+1} L^* \stackrel{\bar \p}{\to} \cdots.
\end{equation}
The cohomology groups of this complex are called the Lie algebroid cohomology groups~$H^k(L)$.
The structure functions $e^{\mu}_{a}$ and $f^a_{bc}$ define
\emph{the anchor map} $L \to TX$ and  \emph{the Lie algebroid bracket}.
The notion of Lie algebroid generalizes the notion of Lie algebra
and the tangent bundle. For a Lie algebra $L$ one can put $e^{\mu}_a=0$ and $f^a_{bc}$ to be
the structure constants of $L$. For the tangent bundle one can put
$e^{\mu}_a$ to be the identity map and $f^a_{bc} = 0$.

For the case when $L$ is the Lie algebroid that describes the $+i$-eigenbundle
of a generalized complex structure on $X$, we will use the supermanifold $\Pi L$
to construct the target space of the $\CalJ$-model.
(In the case of $A$-model, $\Pi L$ is actually the target space).
The algebra $\CalA$ of functions on $\Pi L$ is
(super-)associative, commutative algebra with the differential~$Q$.

The differential $Q$ on $\Pi L$ is not enough to define a quantum
field theory in the BV formalism.
One also needs a BV bracket on the algebra $\CalA = C^{\infty}(\Pi L) =
\Gamma(\Lambda^\bullet L^*)$.\footnote{The differential $Q: \Lambda^{k} L^* \to \Lambda^{k+1} L^*$ is equivalent to the Lie algebroid structure on $L$,
so $Q$ defines a bracket on $\Lambda^\bullet L$, but not on $\Lambda^\bullet
L^*$.}
There is a natural notion to define simultaneously
the compatible differential
and the bracket on $\Pi L$ -- \emph{a Lie bialgebroid} \cite{MR1472888,MR1675117,MR1625610,Roytenberg:thesis}.
A Lie bialgebroid  is a pair $(L,L^*)$ of Lie algebroid $L$ and its dual $L^*$, such that
the differential $Q: \Lambda^k L^* \to \Lambda^{k+1} L^*$ of the Lie algebroid $L$
satisfies the Leibnitz's rule for the Lie algebroid bracket
on $L^*$. This is the case for the pair
$(L,L^*)$ associated with generalized complex structure~\cite{GCS,Roytenberg:thesis,MR1472888}.
The Lie bracket $\{,\}_{L^*}$ on $L^*$ can be extended to $\Lambda^\bullet L^*$.
That equips the algebra $\CalA = C^{\infty}(\Pi L)$ with the odd Poisson bracket of
degree -1. This is the BV bracket. It satisfies the Leibnitz's rule
for the Lie algebroid differential~$\bar \p$. To summarize, a Lie
bialgebroid structure $(L,L^*)$ is equivalent to the structure on
$\CalA = C^{\infty}(\Pi L)$ of a differential odd Poisson
algebra\footnote{Sometimes it is called Gerstenhaber algebra. See
also~\cite{Getzler:1993as,Getzler:1994yd}.}  $(\CalA,\bar \p,[,]_{L^*})$~\cite{MR1625610,MR1675117,MR1906481}.

A (differential) \emph{Batalin-Vilkovisky  algebra} is a special
case of a (differential) odd Poisson  algebra. In a BV
algebra~\cite{Schwarz:1992nx,Schwarz:2000ct,Witten:1990wb}
the Poisson bracket is generated by an odd 2-nilpotent differential operator of the second
order $\Delta$, which is called \emph{BV Laplacian}~\cite{MR1675117,MR1625610}.
It generates \emph{BV bracket} on $\CalA$ according to the formula
\begin{align}
  \label{eq:BV_bracket}
  \{a,b\} = (-1)^{|a|}\Delta (a\wedge b) - \Delta a\wedge b - (-1)^{|a|}a \wedge \Delta b \quad \text{for} \quad a,b \in \CalA.
\end{align}

Such an operator $\Delta$ can be naturally constructed for a  generalized CY manifold
as follows. The spinors of $\Spin(TX,T^*X)$ are identified with all
differential forms $\Omega^\bullet(X)\equiv \Lambda^{\bullet} T^*X$.
The canonical line bundle $U_0$~\cite{GCS,Hitchin:2} is
a \emph{pure spinor} line bundle $U_0$ defined as a subbundle
of those spinors of $\Spin(TX,T^*X)$, which are annihilated by all sections of $L$.
In the Fock space spin representation, the $U_0$ is \emph{the vacuum} line bundle, the
sections of $L$ are lowering operators and the sections of $L^*$
are increasing operators.  The canonical line bundle $U_0$ defines
the alternative grading on $\Lambda^\bullet T^*X$ as follows
\[\Lambda^\bullet T^*X = U_0 \oplus U_1 \oplus \dots \oplus U_{2n},\]
where $U_{k} =  \Lambda^k L^* \otimes U_0$.
It is explained in~\cite{GCS} that
integrability of $L$ is equivalent to the condition $d(\Gamma(U_0)) \subset d(\Gamma(U_1))$.
Then~\cite{GCS} one can define generalized $\p$ and $\bar \p$ operators
on $\Lambda^\bullet ( T^*X)$ in such a way
that $d=\p + \bar \p$ and  $\bar \p: \Gamma(U_k) \to \Gamma(U_{k+1})$
and $\p: \Gamma(U_k) \to \Gamma(U_{k-1})$.
A generalized CY manifold $X$ is defined~\cite{Hitchin:2,GCS} by the condition that
on $X$ exists a nowhere-vanishing closed section $\rho$ of the canonical bundle
$U_0$ -- \emph{`pure spinor' }. (In the case of CY manifold $\rho$ is the
holomorphic $(3,0)$ form. In the case of a symplectic manifold
$\rho$ is $e^{i \omega}$, where $\omega$ is a symplectic
structure.) Using $\rho$, the operators $\p$ and $\bar
\p$ on the differential forms $\Lambda^\bullet T^*X$ can be mapped to the operators $\p$
and $\bar \p$ on sections of $\Lambda^\bullet L^*$ by the formula
\begin{align}
\label{d_bar_d_L}
 \bar \p (\mu \cdot \rho) = \bar \p \mu \cdot \rho, \quad
  \p (\mu \cdot \rho) =  \p \mu \cdot \rho\quad \text{for} \quad
  \mu \in \Gamma(\Lambda^\bullet L^*).
\end{align}
To summarize, a generalized CY structure defines
a differential BV algebra $(\CalA, Q, \Delta)$ where $\CalA = \Gamma(\Lambda^\bullet L^*)$,
$Q =  \bar \p, \Delta = \p$~\cite{MR1675117,MR1472888,GCS,MR1625610,MR1919435,MR1609624}.
The BV bracket $\{,\}$ is defined in terms of $\Delta=\p$ by~\eqref{eq:BV_bracket}.
Since  $\p \bar \p + \bar \p \p=0$, the operator $\bar \p$ is a derivation of the bracket $\{,\}$,
and since $\p^2=0$, the operator $\p$ is a derivation of the bracket $\{,\}$.

To describe quantum field theory in BV formalism one also
needs a measure on the space of fields $\CalA$.
It is called \emph{trace map}  $\tr: \CalA \to \BC$
on the algebra $\CalA$. It satisfies
\begin{align}
  \tr ( (\Delta a)b )  = (-1)^{a} \tr (a \Delta b), \\
  \tr ( (Q a) b ) = (-1)^{a+1} \tr (a\, Q b)
\end{align}

For a  generalized CY manifold, the trace map is
defined by a section $\rho$ of the canonical line
bundle $U_0$. Contracting with $\rho$, one can map
an element $\mu$ of $\CalA = \Gamma(\Lambda^\bullet L^*)$
to a differential form in $\Omega^\bullet (X)$.
This differential form $\mu \cdot \rho$ is also a spinor of $\Spin(TX, T^*X)$.
There is a natural $\Spin(TX,T^*X)$ invariant bilinear
form on $\Omega^{\bullet}(X)$~\cite{GCS,Hitchin:2} with values in $\Omega^{\dim
X}(X)$.
%\begin{align}
%(,): S\otimes S \to \Lambda^{\dim X} T^*X.
%\end{align}
It is given by the wedge product with a certain sign
\begin{align}
(a,b) =  \tilde a  \wedge b,
\end{align}
where $\tilde a = a$ for $\deg a =4k, 4k+1 $, and  $\tilde a= -a$ for $\deg a = 4k+2, 4k+3$.
This bilinear form is symmetric in dimension $4k,4k+1$ and antisymmetric
otherwise.

Using the bilinear form~$(,)$ and the canonical pure spinor $\rho$ we define
\begin{align}
\label{eq:trace_map}
\tr a \equiv \int_X (a \cdot \Omega, \Omega).
\end{align}

\underline{Examples}

1. \underline{The complex case (the B-model).} A generalized complex structure $\CalJ$ that corresponds
to an ordinary complex structure has the following matrix $\CalJ \in \End(TX \oplus
T^*X)$
\begin{align}
\label{eq:J_complex}
\CalJ = \left(\begin{array}{cc}
  -I & 0 \\
  0  & I^T
\end{array}\right),
\end{align}
with $+i$-eigenbundle $L =  TX^{01} \oplus T^*X^{10}$, so that
$L^* = TX^{10} \oplus T^*X^ {01}$. We consider the case of Calabi-Yau,
then the canonical pure spinor is a nowhere vanishing holomorphic $(n,0)$ form~$\rho$.
The algebra $\CalA =C^{\infty} (\Pi L) = \Gamma(\Lambda^\bullet L^*)$  is the
familiar complex~\cite{Witten:1991zz,MR1609624,MR1919435} of the observables of the B-model
$\Lambda^\bullet L^* = \Lambda^\bullet ( TX^{10} \oplus T^*X^{01}) = \Omega^{\bullet}(\Lambda^\bullet TX^{10})$.

The Lie algebroid differential $Q=\bar \p$
is the standard Dolbeault differential $\bar \p:\Omega^{-p,q} \to
\Omega^{-p,q+1}$.
The BV Laplacian $\Delta$ is the
holomorphic divergence $\Delta =\p \cdot : \Omega^{-p,q} \to
\Omega^{-p+1,q}$. The BV bracket on $\CalA$ is generated
by $\Delta = \p$ and can be viewed as the Lie bracket on holomorphic polyvector fields
with coefficients in $(0,q)$-forms. As explained above, the
definition of $\Delta$ depends on the existence of the \emph{pure
spinor}, or, equivalently, measure $\rho$ on the BV manifold. For
the B-model we take the pure spinor to be the holomorphic
$(n,0)$-form.
The Lie algebroid cohomology is the Dolbeault cohomology $H^q_{\bar \p}(\Lambda^p TX^{10})$.

Explicitly, let  $(x^{i},x^{\bar i})$ be  complex coordinates on a CY manifold $X$
and let $(\psi^{i}, \psi_{\bar i})$ be coordinates
in the fibers of $L = TX^{10} \oplus T^*X ^{01}$. Then the algebra $\CalA = C^{\infty}(\Pi L)$ is the
algebra of functions
$f(x^i,x^{\bar i}, \psi_{i}, \psi^{\bar i})$.
If we take local coordinates, where the coefficients of the holomorphic $(n,0)$ form
are constant functions, then the BV operators $Q$ and $\Delta$ have the following
form\footnote{The expression for $Q$ is the same in any coordinates.
However, the expression for $\Delta$ involves holomorphic
$(n,0)$-form $\rho$. For example, if $\rho := \rho \,dx^1 \wedge dx^2 \dots \wedge
dx^n$, then the divergence of the vector field $\mu^{i}$ is
defined as $\Delta \mu = \partial_i \mu^i + \mu^i \p_i log \rho$.
To simplify formulas, in the text below, we will always assume that
the formula for $\Delta$ is written in the local coordinates where $\rho$
is constant. It is easy to recover the general formula, using the definition of $\Delta$ by
means of the Dolbeault differential $\p$ and the isomorphism map $\rho: \Lambda^\bullet(L^*) \simeq \Lambda^\bullet(TX^*)$.}
\begin{align}
Q = \bar \p = \psi^{\bar i} \frac {\p} {\p x^{\bar i}}\\
\Delta = \p = \frac {\p} {\p \psi_i} \frac {\p} {\p x^{i}}.
\end{align}

The trace map on $\CalA$ is familiar from the standard path
integral of the B-model~\cite{Witten:1991zz} $\tr a = \int_X (a \cdot \Omega) \wedge\Omega$.
It can be written as a
Berezian $\rho$ on the superspace $\Pi L$
\begin{align}
\tr a  = \int_{\Pi L} \rho a(x^i,x^{\bar j} ,\psi_{i}, \psi^{\bar j}),
\end{align}
where
\begin{align}
\rho =\Omega_{i_1, \dots, i_n} \frac \p {\p \psi^{i_1}}  \dots \frac {\p}
{\p \psi^{i_n}}  \Omega_{j_1,\dots,j_n}  dx^{j_1}\dots dx^{j_n} dx^{\bar k_1} \dots dx^{\bar k_n}
\frac {\p} {\p \psi_{\bar k_1}} \frac {\p} {\p \psi_{\bar k_n}}.
\end{align}

2. \underline{The symplectic case (the A-model)}.
A generalized complex structure that corresponds to a symplectic
structure $\omega$ has the following matrix $\CalJ \in \End(TX\oplus T^*X)$
\begin{align}
\CalJ = \left(\begin{array}{cc}
\label{eq:J_sympl}
  0 & -\omega^{-1} \\
  \omega & 0
\end{array}\right).
\end{align}
The sections of the $+i$-eigenbundle $L \subset TX \oplus T^*X $ are
given by pairs $(X, -i \omega X)$ where $X \in \Gamma(TX)$ is an arbitrary vector
field. The Lie algebroid $L$ of the A-model is
isomorphic to the tangent bundle $TX$. The Lie algebroid bracket
on $L$ is mapped to the standard Lie bracket on vector fields: one can
check that the restriction of the Courant bracket on $L$ satisfies
\begin{align}
 [ X - i \omega X, Y  - i \omega Y ] = [X, Y] - i \omega [X, Y]
\end{align}
for vector fields $X,Y \in \Gamma(TX)$. The Lie algebroid
differential $Q: \Gamma(\Lambda^{k} L^*)  \to \Gamma(\Lambda^{k+1} L^*)$ is mapped
to the de Rham differential $d: \Omega^{k}(X) \to
\Omega^{k+1}(X)$. The algebra $\CalA = C^{\infty} (\Pi L) = \Gamma (\Lambda^\bullet L^*)$ is
isomorphic to the de Rham complex $\Omega^{\bullet}(X)$.
The Lie algebroid cohomology groups $H^k(L)$ are the de Rham cohomology groups
 $H^k(L) = H^k_{DR}(X)$.
The Lie algebroid bracket on $\Lambda^\bullet L^* \simeq \Lambda^\bullet(T^*X) = \Omega^\bullet(X)$ is the generalization of the Poisson
bracket on functions to the space of differential forms $\Omega^\bullet(X)$.
The BV Laplacian $\Delta: \Omega^{k}(X) \to \Omega^{k-1} (X) $ generates this bracket on
$\Omega^\bullet(X)$~\cite{MR950556,MR1637093}. Explicitly
$\Delta = [\Lambda, d]$, where $\Lambda: \Omega^{k}(X) \to \Omega^{k-2}(X)$ is the operator of
contraction with the Poisson structure $\omega^{-1}$ and $d: \Omega^k(X) \to \Omega^{k+1}(X)$ is the de Rham
differential. The cohomology of $\Delta$ are called
\emph{canonical cohomology} in~\cite{MR950556,MR1637093}.

Explicitly, let $(x^{i},\psi^{i})$ be coordinates in $\Pi TX \simeq \Pi
L$. Then
\begin{align}
 \bar \p = Q = d = \psi^{i} \frac {\p} {\p x^{i}},\\
 \Lambda = \omega^{ij} \frac {\p} {\p \psi^i} \frac {\p} {\p \psi^j}
\end{align}
and $\p = \Delta = \delta = [\Lambda, d]$. In Darboux coordinates,
where $\omega$ is constant, one simply has $ \Delta = \omega^{ij}
\frac {\p} {\p x^i} \frac {\p} {\p \psi^i}$. In other words
$\Delta$ is the symplectic conjugate of $d$.

The trace map on the algebra $\CalA \simeq \Gamma(\Omega(X))$
is defined as the integral of the top degree component
\begin{align}
\tr a  = \int_X a \quad \text{for} \quad a \in \Omega^\bullet(X).
\end{align}

One can also take a dual point of view and consider the
isomorphism $\Pi L \simeq \Pi T^*X$ induced by $\omega$.  Then in the coordinates
$(x^{i}, \psi_{i})$ one has $\Delta = \frac {\p} {\p \psi_i} \frac {\p} {\p
x^j}$ and $\bar \p = Q = \{S, \cdot \}$ with $S = \omega^{ij} \psi_i
\psi_j$.

\underline{Other types of generalized complex structure in the example of K3}

Let $X$ be a generalized CY manifold
equipped with a generalized complex structure  $\CalJ$ and
a canonical pure spinor~$\Omega$~\cite{Hitchin:2,GCS}.
There is a notion of type of generalized complex structure~$\CalJ$~\cite{Hitchin:2,GCS}.
The type of~$\CalJ$
is defined as the codimension of projection
of the associated Lie algebroid $L$ on $TX\otimes \BC$. The sections of $L$
for an ordinary symplectic structure are represented by
$(X^{\mu}, -i\omega_{\mu\nu} X^\nu) \in \Gamma(TX \oplus T^*X)\otimes \BC$,
so $L$ is mapped on $TX$, so  the codimension is $0$, so the type is $0$.
The sections of $L$ of an ordinary complex structure structure are
represented by $(X^{\bar i},\xi_{i}) \in \Gamma(TX^{01},T^*X^{10})$,
so  the codimension is $n = \dim_{\BC} X$.
The type $0$ of an ordinary symplectic structure is the most general type,
and the type $n$ of an ordinary complex structure is the most singular type.
Under deformation, the type of complex structure changes by even numbers.
There is well defined notion of chirality of a given complex structure.

A canonical pure spinor of $Spin(TX,T^*X)$, which is a spinor annihilated by all sections of
$L$, can be represented by a differential form
of mixed degree in $\Omega^{odd/even}(X)$.
The type is the degree of the lowest component of this differential form.
For an ordinary symplectic structure  $\Omega = e^{i\omega}$.
For an ordinary complex structure $\Omega$ is a holomorphic $(n,0)$ form.

Since the A-model generalized CY structure is of type $0$,
and the B-model generalized CY structure is of type $\dim_{\BC} X$,
we see that if $\dim_{\BC} X$ is odd, the A-model and B-model are of different chiralities,
therefore, a generalized $\CalJ$-model in  \emph{odd} complex dimension
\emph{cannot} interpolate between the A-model and the B-model.
Still it~\emph{might} interpolate if $\dim_{\BC} X$ is even. For example, we can take
the B-model and deform it by a holomorphic Poisson bivector $\beta$. This
deformation on the other hand can be viewed as B-field deformation of the A-model.

For example, consider the complex dimension two, and take
$\Omega_A = e^{b+ i\omega}$ and $\Omega_B = \sigma$,
where $\sigma$ is the holomorphic $(2,0)$ form.
Now consider a deformation of B-model by the holomorphic bivector $\beta^{ij}$ such that $\beta\cdot \sigma =1$.
Under this deformation $\Omega_B \to e^{\beta}\sigma = 1 + \sigma$.
On the other hand take the symplectic structure of the A-model $\omega = \I \sigma$,
and then consider deformation of this A-model with $\Omega_A = e^{i \omega}$
by the B-field $b = \R \sigma$. Under such a  deformation
$\Omega_A \to e^{b+i\omega } = e^{\sigma} = 1 + \sigma$, since $\sigma \wedge \sigma = 0$.
Therefore, a holomorphic Poisson bivector $\beta$-field deformation of the B-model
is the B-field deformation of the $A$-model\footnote{
In~\cite{Kapustin:2005vs} it argued that the relation holds for D-branes.
More precisely, let $X$ is a complex manifold equipped with holomorphic $(2,0)$ form $\sigma=b+i\omega$.
Then the $b$-field transformation of the A-model $(X,\omega)$ is
equivalent to the $\beta$-transformation by $\beta = \sigma^{-1}$ of the
$B$-model. In terms of matrices of generalized complex structure for $b$-transformation of symplectic structure
$\omega$ we have
$
\CalJ = \left(\begin{array}{cc}
  \omega^{-1} b & -\omega^{-1} \\
  \omega + b \omega^{-1} b & -b \omega^{-1}
\end{array}\right)
$. This is actually a $\beta$-transformation of the complex structure $I = \omega^{-1}
b$. (One has $\omega + b \omega^{-1} b = 0$ as the condition for the existence of space-filling coisotropic
branes in the $A$-model~\cite{Kapustin:2001ij}.)}.

In complex dimension two, we can take, for example, $X$ to be the $K3$ surface~\cite{MR2115675}.
Then moduli space of generalized complex structures is of even type
(which is the type of the A-model and the B-model in two dimensions), is parameterized
by the Grassmanian $Gr_{2}(\BR^{24})$
 of real 2-planes in the  real 24-dimensional space $H^{\bullet}(K3)=H^{even}(K3)\simeq
 \BR^{24}$.
A 2-plane is spanned by real and imaginary part of the canonical pure spinor $\Omega$.
The physical $N=(2,2)$ CFT with $K3$-surface target space requires  existence
of \emph{two} generalized orthogonal complex structures $(\CalJ,\CalJ')$
represented by \emph{two} orthogonal pure spinors $\Omega_1, \Omega_2$.
Each of these spinors spans a 2-plane in $H^{\bullet}(K3)$,
and these planes must be orthogonal. Therefore, the moduli space of $N=(2,2)$ CFT is locally
parameterized by the Grassmanian $Gr_{2,2}(\BR^{24})$
of two orthogonal 2-planes in $\BR^{24}$.
The moduli space of $N=(4,4)$ CFT's is locally  the space  of 4-planes  $Gr_4(\BR^{24})$.
The moduli space of $N=(2,2)$ CFT's fibers over the moduli space of $N=(4,4)$ CFT's. The fibers
$S^2 \times S^2$ parametrize decomposition of a 4-plane into two orthogonal 2-planes~\cite{Aspinwall:1994rg,Aspinwall:1996mn}.

The topological twisting~\cite{Witten:1991zz} makes theory
to depend only on a half of this $\CalN=(2,2)$ CFT structure.
That is, the twisted theory couples only to $\CalJ$ and
does not couple to $\CalJ'$.
Thus, we see that in the case of $K3$ surface, the present
construction is in agreement with the traditional considerations~\cite{Aspinwall:1994rg,Aspinwall:1996mn} and~\cite{Kapustin:2003sg}.

\section{The two-dimensional topological $\CalJ$-model\label{2Dtheory}}

%Now let us construct a topological model by the data $(A,\Delta,\dL,tr)$.
%We will see that, given this data, a target space formulation is more natural,
%or at least it is more useful to compute all correlation functions in a large volume limit,
%when we can neglect instanton corrections. Still, the worldsheet formulation will be useful to compute a deformation of the product
%in the algebra of the open string observables on a disk for a $(-2,0)$-deformation of the B-model.

The data $(X,Q,\Delta,\rho)$ associated with a generalized CY
manifold $X$ allow us to construct a topological $\CalJ$-model
using the AKSZ method~\cite{MR1432574}.

\subsection{The AKSZ construction of sigma-model for $\Maps(\hat \Sigma, M)$ for a target space $M$ with
a $PQ$-structure}

A $PQ$-manifold is a supermanifold equipped with an odd symplectic structure $\omega$
and a Hamiltonian vector field $Q$. In~\cite{MR1432574} a topological sigma-model
was constructed for any such target space~$M$.

Let us review the key properties of a $PQ$-target space.
The symplectic structure defines the odd Poisson bracket $\{\cdot,\cdot\}$.
Since $\omega$ is $Q$-invariant
\begin{equation}
  \label{eq:PQ_compatible1}
  L_Q \omega = (di_Q + i_Qd) \omega = d(i_Q \omega) =0,
\end{equation}
the one-form $i_Q \omega$ is closed. We consider the case when $i_Q \omega$ is exact,
so there is a function $S$ such that  $i_Q \omega = dS$. Such a
function on $X$ is called~\emph{Hamiltonian function} for the
vector field $Q$. For any function
$f \in C^\infty(M)$ its Lie derivative $L_Q f$ can be computed
as a bracket with the Hamiltonian function $S$ associated with $Q$
\begin{equation}
  \label{eq:LQF}
  L_Q f = \{S,f\}.
\end{equation}
The homological property $Q^2=0$ of the $Q$-structure can be written as
the \emph{BV classical master equation}~\cite{Schwarz:1992nx,Schwarz:2000ct,Witten:1990wb}
\begin{equation}
  \label{eq:SS=0}
  \{S,S\}=0.
\end{equation}

A canonical example of a $Q$-manifold is a tangent bundle with
parity reversed on the fibers $\Pi TX$. In coordinates $(x^{\mu},
\psi^{\mu})$ one has $Q = \psi^{\mu} \p_{\mu}$. As discussed in
the previous section, the tangent bundle is an example of a Lie algebroid.
The total space $\Pi L$ of any Lie algebroid is a particular case of
$Q$-manifold with the vector field $Q$ of degree 1.

A canonical example of a $P$-manifold is a cotangent bundle with
parity reversed on the fibers $\Pi T^*X$. In coordinates $(x^{\mu},
\pi_{\mu})$ the canonical symplectic form is
$\omega(\delta x, \delta \pi) = \delta x^{\mu} \delta \pi_{\mu}$.

A $PQ$-manifold $M$ can be constructed starting from any $Q$-manifold $N$ as $M = \Pi T^*N$.
If $(x^\mu, \pi_\mu)$ are coordinates on $\Pi T^*N$, then for any vector field
$Q=v^{\mu} \frac {\p} {\p x^{\mu}}$  on $N$ there is a Hamiltonian function $S = v^{\mu} \pi_\mu$ on $M$
which generates $Q$ on $N$.

Let $\Sigma$ be a two-dimensional bosonic worldsheet.
We extend it by fermionic directions and consider the
$Q$-supermanifold $\hat \Sigma = \Pi TX$ with a canonical measure.
That is we take the total space $\Pi T\Sigma$ of the tangent bundle of $\Sigma$ with
parity reversed on the
fibers. We equip $\hat \Sigma = \Pi T \Sigma$ with the standard Berezian measure. If $(\sigma,\theta)$
are coordinates on $\hat \Sigma$, the standard
measure is $\rho = d \sigma^1 d\sigma^2
\frac {\p} {\p \theta^1} \frac{ \p} {\p \theta^2}$. Functions on
$\hat \Sigma = \Pi T \Sigma$ are differential forms
$\Gamma(\Omega^\bullet(\Sigma))$. The product of functions on $\Pi T \Sigma$ is
the wedge product of differential forms on $\Sigma$. For any function $f$ on $\Pi T
\Sigma$, which is a collection of differential forms $f^{(i)}$ of all
degrees $f = f^{(0)}+ f^{(1)} + f^{(2)}$, one can take the
highest component~$f^{(2)}$ and integrate
over~$\Sigma$, so we define the integral $\int f := \int_{\Sigma} f^{(2)}$.
Equivalently, it is the integral over $\hat \Sigma$ with the standard Berezian measure
\begin{align}
\label{eq:def_int_ptsigma}
\int f : = \int _{\hat \Sigma} \rho f.
\end{align}
In the following formulas the standard Berezian measure
$\rho = d \sigma^1 d\sigma^2 \frac {\p} {\p \theta^1} \frac{ \p} {\p
\theta^2}$ on the worldsheet will be
omitted under the sign of integral. The $Q$-structure $Q_{\Sigma} = \theta^{i} \frac {\p} {\p
\sigma^{i}}$ on $\hat \Sigma$
is the standard de Rham differential $d$ on $\Sigma$.

The idea of~\cite{MR1432574} is to construct the BV structure on the
space $\Maps(\hat \Sigma, M)$ using the $Q$-structure $Q_{\Sigma}$ on the
worldsheet, the $Q$-structure $Q_{M}$ on the target space
generated by the Hamiltonian function $S_{M}$, the odd symplectic
structure on the target space $\omega$ and the integral on the worldsheet $\int_{\hat \Sigma}$.

The space $\Maps(\hat \Sigma, M)$ is the BV phase
space of the model.
The odd symplectic structure $\hat \omega$ on $\Maps(\hat \Sigma, M)$
is defined by the integral of the pullback of the odd symplectic structure $\omega$
from the target space.
For variations $\delta \phi_1, \delta \phi_2$ of a map $\phi \in \Maps (\hat \Sigma, M)$
we define the value of $\hat \omega$ on $\delta \phi_1, \delta \phi_2$ to be
the integral over $\hat \Sigma$
\begin{align}
  \label{eq:def_P_on_sigma2M}
  \tilde \omega(\delta \phi_1,\delta \phi_2) =
  \int_{\hat \Sigma} \, \omega(\delta \phi_1, \delta \phi_2).
\end{align}

As for the $Q$-structure on $\Maps(\hat \Sigma, M)$ we can take
\begin{align}
  \label{eq:def_Q_on_sigma2M}
  Q  = Q_{\Sigma} + Q_{M}.
\end{align}
Any $Q$-structure generates diffeomorphism. There is a natural action of $Diff(\hat \Sigma)$
and $Diff(M)$ on the space $\Maps(\hat \Sigma, M)$.
The $Q$-structure~\eqref{eq:def_Q_on_sigma2M} corresponds to
 composition of two
diffeomorphisms~\cite{MR1432574,MR1854134}.
For any such $Q$ there is  a corresponding Hamiltonian function
$S$. The function $S$ is the called~\emph{BV master action}~\cite{Batalin:1981jr,Batalin:1984jr}.

To write down $S_{\Sigma}$, it is convenient to represent the
odd symplectic target space $M$ as $\Pi T^* N$ for some manifold
$N$. Let $x^{\mu}$ be coordinates on $N$ and $p_{\mu}$ be
coordinates on the fiber of $\Pi T^*N$. Then $x^{\mu}$ and
$p_{\mu}$ are canonically conjugate fields on $\hat \Sigma$.
The Hamiltonian function $S_{\Sigma}$, which generates the de Rham differential
$Q_{\Sigma} = d$, is
\begin{align}
  \label{eq:S_sigma}
  S_\Sigma[p,x] = \int_{\hat \Sigma } p_{\mu} dx^{\mu}.
\end{align}

The Hamiltonian function $S_{M}$ for the structure $Q_{M}$ is the integral over $\hat \Sigma$ of the pullback of $S_{M}$
\begin{align}
  \label{eq:tilde_S_M}
  S_{M}[p,x]  = \int_{\hat \Sigma} S_{M} (p,x).
\end{align}

The total BV master action of the topological model is $S = S_{\Sigma} +
S_{Q}$
\begin{align}
  \label{eq:fullBVaction}
  S[p,x]  = \int_{\hat \Sigma} p_{\mu} dx^{\mu} + S_{M} (p,x).
\end{align}

Let us check that $S = S_{\Sigma} + S_{M}$ satisfies the BV classical master equation $\{S,S\} = 0$.
That is equivalent to $\{S_{\Sigma},S_{\Sigma}\} = 0$,
$\{S_{M}, S_M\} = 0$ and $\{S_{\Sigma}, S_M\}=0$. The first two
equations are satisfied because $Q_{\Sigma}^2 = 0$ and
$Q_{M}^2=0$. The third equation
\[
\{S_{\Sigma}, S_M\} = Q_{\Sigma} S_{M} =
\int_{\hat \Sigma} d S_{M}= \oint_{\p \Sigma} S_{M}\]
is automatically true for a closed surface. If $\Sigma$ has a boundary,
then in order to satisfy the BV classical master equation
for $S$, we need to impose the boundary conditions $S_{M}|_{\p \Sigma} = 0$.
The model~\label{eq:fullBVaction} is a
BV version of
the Poisson
sigma-model~\cite{Schaller:1994es,Schaller:1994uj,Kotov:2004wz,MR1854134,Ikeda:2002qx,
Ikeda:1993fh,Ikeda:2004cm,Ikeda:2001fq}.
The functional integral is supposed to be taken over a Lagrangian
submanifold in the BV phase
space~\cite{Batalin:1981jr,Batalin:1984jr,Schwarz:1992nx,Schwarz:2000ct}
of fields $(p_{\mu},x^{\mu})$ on  $\hat \Sigma$.

The space of functions on a $PQ$-manifold $M$ with a measure is
a differential BV algebra. More precisely~\cite{Schwarz:1992nx,MR1675117,MR1625610}
the algebra $C^\infty(M)$ for a general $PQ$-manifold is
a differential odd Poisson algebra. However it is not always a BV algebra.
It is a BV algebra if $M$ is equipped with a generator of the Poisson bracket --
BV Laplacian $\Delta$, which can be constructed by a measure.
A BV Laplacian is an odd differential 2-nilpotent operator $\Delta$ of the second
order generating the Poisson bracket.
The canonical example  $\Pi T^*N$ of a
$P$-manifold with coordinates $(x^{\mu},p_{\mu})$
does have such an operator $\Delta$. It has explicit
form $\Delta = \frac {\p} {\p p_{\mu}} \frac {\p} {\p x^{\mu}}$.

\subsection{Construction of target space with $PQ$-structure for a
generalized CY manifold}
We reviewed the AKSZ procedure~\cite{MR1432574}
of constructing a topological sigma-model for a target space $M$
with $PQ$-structure. Now we need to construct such a space $M$ starting
from a generalized CY manifold $X$.

As explained in section~\ref{BValgebra},
there is a differential BV algebra $(\CalA,Q,\Delta,\rho)$ of functions on $N=\Pi L$, which
is  the total space of the Lie algebroid $L$ with parity reversed on the fibers.
In other words, $N$ is equipped with the odd Poisson bracket
$\{,\}$ and the $Q$-structure $\bar \p$. Recall that a $PQ$-manifold
is a supermanifold equipped with an odd symplectic structure $\omega$
and a Hamiltonian vector field $Q$. That means that $N=\Pi L$ is a
$PQ$-manifold if the Poisson structure is an inverse of some symplectic structure $\omega$ and
$Q$ is generated by some Hamiltonian function $S$.

In the AKSZ approach, the symplectic structure is required to
construct the BV master action and its gauge fixed version. It is
possible, however, that one could define amplitudes of the generalized topological sigma-model
starting from any differential BV algebra $(\CalA,Q,\Delta,\rho)$,
without assumption, that the Poisson structure generated by
$\Delta$, is invertible. As we will see later, this is indeed the case
for tree level amplitudes in the large volume limit.
Using only BV algebra  $(\CalA,Q,\Delta,\rho)$, it is
possible\footnote{with some
additional mild assumption similar to the $\bar\p \p$-lemma, see~\cite{Losev:0506039}} to define closed
topological string filed theory, which
generalizes the Kodaira-Spencer theory
of~\cite{Bershadsky:1993cx}. It definitely works in the genus zero
and in the sector of topologically trivial maps, but it is not yet
clear whether the BV algebra $(\CalA,Q,\Delta,\rho)$ completely
defines the full theory, however there are some
indications~\cite{Losev:0506039}.

So far we will take more pragmatic approach and will try to
reduce the problem to the AKSZ framework. The problem with the manifold $N= \Pi L$ is
that the Poisson structure on it is not always invertible. For example,
it is invertible in the case of the A-model,
where $\Pi L \simeq \Pi T^*X$. However, it is
\emph{not} invertible in the case of the B-model, where $\Pi L =
TX^{01} \oplus T^*X^{10}$. Indeed, in the coordinates $(x^{i}, x^{\bar i}, \psi^{i}, \psi_{i})$
the odd Poisson bracket is generated by
the BV Laplacian $\Delta = \frac {\p} {\p x^i} \frac {\p} {\p
\psi_i}$. We can see that $x^{i}$ and $\psi_i$ are conjugate fields to each other, but
$x^{\bar i}$ and $\psi^{\bar i}$ do not have conjugates. The
Poisson structure degenerates on the fields $x^{\bar i}$ and
$\psi^{\bar i}$.
In this case it is impossible to find a Hamiltonian function~$S$, such that it
generates the operator $Q = \psi^i \frac {\p} {\p x^{i}}$ via
the Poisson bracket as $Q = \{S, \cdot\}$. The AKSZ construction
cannot be used directly.

One way to solve this problem is to extend the set of fields $\Pi L$ by
auxilary fields and to make the Poisson structure nondegenerate.

To construct $M$, let us recall, that any Poisson manifold $N$
in a vicinity of a regular point is a symplectic fibration. Let $K$ be the space of
symplectic leaves. The Poisson bracket vanishes for functions that
are constant along the leaves, i.e. for functions that depend only
on $K$. Explicitly, near regular point of $N$,
 one can pick up the local coordinates $(p_i,q^j,z^a)$ on $N$ in such a
 way\footnote{By the Darboux theorem for Poisson manifolds},  that the Poisson structure has the standard form:
 the only non-vanishing bracket is $\{p_i,q^j\} = \delta_i^j$.
 Here $(p_i,q^j)$ are the local coordinates on the symplectic fibers,
 and ${z^a}$ are the local coordinates on the base $K$, i.e. $\{z^a\}$
 are Casimir functions on $N$. Let us
 introduce the new fields $z_a^*$, which are conjugate to $z^a$,
 and then consider the manifold $M$ with the local coordinates $(p_i,q^j,z^a,z_a^*)$.
The manifold $M$ is a symplectic manifold, moreover,
the projection $\pi: M \to N$ is a Poisson map.\footnote{A map
$\pi: M \to N$ is called Poisson if
$\pi^* ( \{f,g\}_N ) = \{\pi^*(f), \pi^*(g)\}_M$.}
The manifold $M$ is a symplectic realization of
$N$~\cite{MR1747916,MR866024,MR854594}, moreover $ N \hookrightarrow M$ is
a leaf-symplectic embedding~\cite{Bojowald:2001ae}.
This local construction might fail to work at non-regular points of $N$,
where the rank of the Poisson structure is not constant, and there
also could be global obstructions~\cite{Bojowald:2001ae}. We leave
these two important and interesting questions for the future
study and assume in the present work that such symplectic realization
$M$ is well defined globally.

The vector field $Q$ on $N$ can be generated
by a Hamiltonian function $S$ on $M$, i.e. for functions on $N$ one has
$Q = \{S, \cdot\}$. The function $S$ on $M$ defines the vector field $\tilde Q$
on $M$. The cohomology of $\tilde Q$ on $M$ is isomorphic to the cohomology of
$Q$ on $N$, so the physical states in the algebra $C^\infty(M)$ are the same
as in the algebra $C^\infty(N)$.

\underline{Example}. The B-model.
    Let $X$ be a CY manifold, $\dim_\BC X = n$. Consider the
generalized complex structure $\CalJ$ corresponding to the complex
structure on $X$. The Lie algebroid of the $+i$-eigenbundle $\CalJ$
is $L = TX^{01} \oplus T^*X^{10}$. The total space of $L$ with
fermionic fibers is the supermanifold $N = \Pi L$.
The BV algebra of observables is the algebra of functions $\CalA = C^{\infty}(\Pi TX^{01}
\oplus \Pi T^*X^{10})$, equivalently the algebra of sections of
$\Lambda^k(L^*)=\oplus_{p+q=k} \Lambda^{p}(TX^{10}) \otimes \Lambda^{q} (T^*X^{01})$.
We pick up local coordinates $(x^{i},x^{\bar i}, \psi_{i}, \psi^{\bar i})$ on $\Pi L$,
then functions on $\Pi L$ are expanded as $f_{\bar i_1,\dots \bar i_q}^{i_1\dots i_p}
\psi^{\bar i_1} \dots  \psi^{\bar i_q}  \psi_{i_1} \dots
\psi_{i_p}$.
The Lie algebroid differential $Q$ is the Dolbeault differential $\bar \p$
\begin{align}
 \label{eq:Q_B_model}
  Q \equiv \bar \p  = \psi^{\bar i} \frac {\p} {\p x^{\bar i}}.
\end{align}
The $Q$-cohomology of the algebra $\CalA$ is the familiar space of
physical observables of the
B-model~\cite{Witten:1991zz,MR1609624}.

The BV Laplacian  $\Delta$ is the holomorphic divergence $\p$
on holomorphic vector fields $\Lambda^p(TX^{10})\otimes \Lambda^q(T^*X^{01})$
that can be mapped by the holomorphic $(n,0)$ form
to the Dolbeault differential $\p$ on the differential
forms $\Lambda^{n-p}(T^*X^{10})\otimes \Lambda^q(T^*X^{01})$.
In coordinates, where the holomorphic $(n,0)$ form is constant,
the operator $\Delta$ can be written as
\begin{align}
\label{eq:Delta}
 \Delta \equiv \p = \frac {\p} {\p \psi_i} \frac{ \p} {\p x^{i}}.
\end{align}
The Poisson bracket, generated as in~\eqref{eq:BV_bracket} by $\Delta$,
is equivalent to the Lie bracket on holomorphic polyvector fields.

Looking at~\eqref{eq:Delta} we see that the Poisson structure is
non-degenerate on the space $(x^{i},\psi_i)$, but it is degenerate
on the space  $(x^{\bar i}, \psi^{\bar i})$.
In other words, $x^{i}$ and $\psi_{i}$ are antifields to each other,
but $x^{\bar i}$ and $\psi^{\bar i}$ are missing their antifields.
So we extend the bundle $\Pi L$ by additional fields canonically conjugate to $x^{\bar i}$ and $\psi^{\bar i}$.
Let us call them $x_{\bar i}^*$ and $\psi_{\bar i}^*$.
To summarize, $M$ is a direct sum of the vector bundles
 $ \Pi TX^{01} \oplus \Pi T^*X^{10} \oplus \Pi T^*X^{01} \oplus
 T^*X^{01}$ over the base $X$.
If local coordinates on $X$ are $(x^i, x^{\bar i})$, then the local coordinates on the fibers
of $M$ are correspondingly $(\psi^{\bar i}, \psi_{i}, x_{\bar i}^*, \psi_{\bar
i}^*)$. The total space of fields of the BV sigma-model is
$(x^i, x^{\bar i}, \psi^{\bar i}, \psi_{i}, x_{\bar i}^*, \psi_{\bar
i}^*)$. The present example of the construction of the B-model differs from the
AKSZ~\cite{MR1432574}~\footnote{In their construction the authors start from 8 fields instead of 6.},
but agrees with~\cite{Hofman:2002cw}.
The odd symplectic structure $\omega$ on~$M$ has a local form
\begin{align}
\label{eq:sympl_B_model}
 \omega = \de x^i \de \psi_i + \de x^{\bar i} \de x_{\bar i}^* + \de \psi^{\bar i} \de \psi_{\bar i}^*.
\end{align}
The Poisson bracket generated by~\eqref{eq:sympl_B_model} and restricted for functions on $\Pi L$
is the same as the Poisson bracket generated by~\eqref{eq:Delta}.
The Hamiltonian function $S$ for the vector field
$Q$~\eqref{eq:Q_B_model} is quadratic in the coordinates of the
fibers of $M$
\begin{align}
 S_{M} = \psi^{\bar i} x_{\bar i}^*.
\end{align}
This function $S_M$ gives us the master action of the B-model.
More precisely, as explained above, the action
of the B-model is obtained
by pulling back the function $S_M$ to a super Riemann surface
$\hat \Sigma$ by means of $\Maps(\hat \Sigma, M)$. One can also
add the term $S_{\Sigma}$ for the de Rham differential $d$ on
$\Sigma$. Finally the BV master action of the B-model is
\begin{align}
S_{B} = \int_{\hat \Sigma} \psi_i dx^i + \psi^{\bar i} x_{\bar
i}^*.
\end{align}

The fields $x^{i}$ and $x^{\bar i}$ are treated as independent variables,
the complex conjugate condition is relaxed.
In the classical limit of the BV formalism one usually considers
deformations of the master action
satisfying the condition $\{S,S\}=0$ and keeps the same measure and the odd Poisson
bracket on the space of antifields.
In the language of the B-model, that means to deform the operator $Q = \bar
\p$, but to keep fixed the operator $\Delta = \p$ and the holomorphic $(n,0)$-form of the base point.
This issue is related to the
holomorphic anomaly of the
B-model~\cite{Witten:1991zz,Bershadsky:1993cx}.
Given a base point in the moduli space of CY structures on $X$, the nearby
complex structures can be parameterized as
deformations of the operator $Q = \bar \p$, and the corresponding
observables can be identified with deformations of the master
action $S$. However, in this approach, in the definition of the BV path
integral, one keeps fixed the measure
and the odd Poisson bracket defined by the base point. If the base point
is changed, then the definition of the Laplacian, the odd Poisson
bracket, and, thus, the measure in the functional integral in the
BV phase space is changed. That is the origin of the holomorphic
anomaly~\cite{Witten:1991zz,Bershadsky:1993cx}. There is a
dependence of the partition function on the base point, since the base point defines the measure in
the BV path integral. It is possible, that a background
independent model could be formulated along the lines of~\cite{Gerasimov:2004yx,Dijkgraaf:2004te,
Hitchin,Hitchin:2,Nekrasov:2004vv,Pestun:2005rp}.

In the present formulation, it is convenient
to study the generalized topological $\CalJ$-model
with respect to a simple base point on the moduli space of generalized complex structures.
In some cases it might be possible to take the ordinary B-model as a base point.
Then topological $\CalJ$-model is nothing else but a finite deformation
of the B-model in generalized complex directions.
At the classical level it was studied in details in~\cite{MR1609624}.

We will consider an explicit example of a generalized complex
structure $\CalJ = (I, \beta)$
described by means of an ordinary complex structure $I$ and the holomorphic
Poisson bivector $\beta$. The type of such generalized complex structure
jumps where $\beta$ vanishes: the description of the manifold $N = \Pi L$ is
complicated. However, if we take the base point
to be the ordinary complex structure~$I$, then the full $\CalJ$-model can be
described by means of deformations of the master action $S$ only.
Explicitly $S_{\CalJ} = S_{I} + \beta^{ij} \psi_i \psi_j$.
That describes generalized deformation of the homological vector
field $Q =\bar \p$ in the frame where the odd Poisson bracket is
fixed to be the same as defined by $I$.
The target space and the measure of such $\CalJ$-model is the same as of the ordinary B-model.
The difference is only in the master action.

We will leave the study of the dependence of the $\CalJ$-model
on the base point for the future work. That should bring a
generalization of the holomorphic anomaly equation. Physically,
the holomorphic anomaly equation was explained in~\cite{Witten:1993ed}
by a change of holomorphic polarization in $H^3(X,\BC)$ with a change of the base point.
In the BV formulation that corresponds
to a change of the definition of the BV Laplacian $\Delta$
and the BV bracket, if one changes the base point.

\underline{Example}. The A-model.
Consider a generalized complex structure $\CalJ$
defined by a non-degenerate symplectic structure $\omega$.
The symplectic structure provides the isomorphism between the
tangent $TX$ and cotangent $T^*X$ bundle  of $X$. The Lie algebroid $L$ is isomorphic to each of them,
and each of $TX$ or $T^*X$ can be used as a model for the target space
of the A-model. Let us take the $\Pi L \simeq T^*X$ and consider the canonical
coordinates on it~$(x^{i},p_{i})$. The odd symplectic
structure is simply $\omega = \delta x^{i} \delta p_{i}$. The
$Q$-structure is generated by the Hamiltonian function
$S_{M} = \omega^{ij}p_{i} p_{j}$. For the A-model the BV Poisson
structure is nondegenerate on $N = \Pi L$, therefore no completion
of $N$ is required. The target space of $\CalJ$-model is simply $M = N$.
The BV action of the A-model is the same as of the Poisson
sigma-model~\cite{Ikeda:1993fh,Schaller:1994uj,MR1432574,MR2023844}
\begin{align}
  S  = \int_{\hat \Sigma} p_i dx^i  + \omega^{ij} p_i p_j.
\end{align}

\subsection{The BV gauge fixing.}

The quantization in the BV formalism~\cite{Batalin:1984jr,Batalin:1981jr,Schwarz:1992nx,Witten:1990wb}
is done by taking  path
integral over a Lagrangian submanifold $L$ in the BV phase space of
the theory
\begin{align}
    Z = \int_L  e^{-\frac 1 \hbar S }.
\end{align}
The partition function does not depend on a deformation of the
Lagrangian submanifold $L$. Indeed, let us represent $M$ as $T^*L$ with the coordinates $(x^i, p_i)$.
The Lagrangian manifold $L$ is the zero section of $T^*L$.
A Lagrangian deformation is defined by a function
$\Psi$ on $L$ as $p_i = \frac {\p} {\p x^i} \Psi$.
A deformation of the integral  $\int_L f$ of any harmonic
function $\Delta f = 0$ vanishes
\begin{align}
  \int_{L+\delta L} f - \int_{L} f = \int_L \frac{\p f} {\p p_i}
  p_i  = \int_L \frac {\p f} {\p p_i} \frac {\p} {\p x^i} \Psi =
  \int_L \Psi \frac{\p} {\p x^i} \frac {\p f} {\p p_i}   = \int_L
  \Psi \Delta f = 0.
\end{align}

A restriction of the BV master action $S$ to a Lagrangian
submanifold is called \emph{gauge fixing}. Locally, a deformation
of a Lagrangian submanifold $L$ can be described by a function $\Psi$ on $L$ which is called
the \emph{gauge fixing fermion}.

To show the idea let us consider the BV gauge fixing
of the ordinary B-model defined on a Kahler manifold $X$.
The Kahler metric of $X$ will be used to conveniently describe a
Lagrangian submanifold in the BV phase space of fields on $\hat
\Sigma$.
In a general case, the $\CalJ$-model might be gauge fixed
using another  generalized complex structure $\CalJ'$ which commutes with
$\CalJ$. In the case of the B-model such generalized complex
structure $\CalJ'$ is a compatible symplectic structure. In the
case of $A$-model such generalized complex structure $\CalJ'$ is
a compatible complex structure.
The geometry of $(\CalJ,\CalJ')$ is a generalized Kahler geometry~\cite{GCS}.
In~\cite{Kapustin:2003sg,Kapustin:2004gv,Kapustin:2005uy} it was suggested
to make a generalized topological B-model by twisting a certain sigma-model.
This sigma-model a priori depends on both $(\CalJ,\CalJ')$,
but dependence on $\CalJ'$ disappears after a twist.
As Gualtieri showed~\cite{GCS}  the generalized Kahler geometry
$(\CalI,\CalJ)$ is equivalent to the data $(g,b,I_+,I_-)$ with
certain compatibility conditions, where $g$ is an ordinary metric,
B is a two-form, $I_+$ and $I_-$ are the ordinary complex
structures. This geometry was encountered in a study of $\CalN=(2,2)$ CFT
 in~\cite{Gates:1984nk}, and recently studied in
 works~\cite{Zabzine:2005qf,Lindstrom:2004hi,Lindstrom:2004iw,Zabzine:2004dp,Lindstrom:2004eh,
Zucchini:2005rh,Zucchini:2004ta,Zucchini:2005cq,Fidanza:2003zi,Grana:2004bg,Chuang:2005qd,Lindstrom:2005zr,Hsu:2006vw,Bredthauer:2006hf}.

From the example of  the A-model (Gromov-Witten
invariants) which exists on any almost Kahler manifold, where $\CalJ$ is a symplectic structure,
and $\CalJ'$ is almost complex structure, we
know that $\CalJ'$ does not have to be integrable. Therefore, an existence of
integrable $\CalJ'$ that together with $\CalJ$ makes a generalized Kahler
geometry, seems not to be generally required for a definition of the topological model.
However, in the case when integrable $\CalJ'$ is used for the gauge
fixing of the topological $\CalJ$-model, one should recover
a twisted version of sigma-model with generalized Kahler space like
in~\cite{Lindstrom:2005zr}.

For an illustration let us consider the ordinary B-model.
The BV master action is written as
\begin{align}
S_{B} = \int_{\hat \Sigma} \mu ( \psi_i dx^i + \psi^{\bar i} x_{\bar i}^*)
\end{align}
in terms of the fields  $(x^i, x^{\bar i}, \psi^{\bar i}, \psi_{i}, x_{\bar i}^*, \psi_{\bar i}^*)$
in the phase space $\Maps(\hat \Sigma, M)$, where $M$ was described  above.
 A field on $\hat \Sigma$ is a collection of differential forms
of all degrees on $\Sigma$. The symplectic pairing is given by the wedge product
and integral over
$\Sigma$ in each of the pairs
$(x^i,\psi_i), (x^{\bar i},x_{\bar i}^*), (\psi^{\bar i}, \psi_{\bar i}^*)$.
As in~\cite{MR1432574}, we will choose a Lagrangian submanifold in two steps.
First, let us algebraically choose one half of the fields in such a way
that the symplectic form vanishes on them. This set of fields will be called antifields
$\Phi^*$. The submanifold of the phase space, where all antifields vanish, is a Lagrangian
submanifold $L: \Phi^* = 0$.  Such naive choice of $L$ gives
a degenerate physical action. To get a nondegenerate physical action
we can deform $L$ by a gauge fixing fermion to get
\begin{align}
\label{eq:LagPsi}
  L_{\Psi} :
 \Phi^* = \frac {\p \Psi} {\p \Phi}.
\end{align}

For illustration purposes let us write the usual derivatives
instead of covariant,  which one has to use in a non-flat case.

The full set of fields represented by differential forms on
$\Sigma$ is written in this diagram.\footnote{Hopefully, the mixing of notation $x^i$ for a field on $\hat \Sigma$ and its zero degree component on $\Sigma$ will not cause confusion,
but will be clear from the context.
If we need to distinguish a field $\phi$ on $\hat \Sigma$ from its zero component,
then we will use the notation $\hat \phi$ for a field on $\hat \Sigma$}
\begin{align}
\label{eq:myfields}
  \begin{array}{cccc}
    x^i    & x^i_z  & x^i_{\bar z} & x^i_{z \bar z}  \\
    \psi_i & \psi_{iz} & \psi_{i\bar z} & \psi_{i z \bar z} \\
    x^{\bar i} & x^{\bar i}_{z} & x^{\bar i}_{\bar z} & x^{\bar i}_{z \bar z}\\
    x_{\bar i}^* & x_{\bar i z}^* & x_{\bar i \bar z}^* & x_{\bar i z \bar z}^* \\
    \psi^{\bar i} & \psi^{\bar i}_z & \psi^{\bar i}_{\bar z} & \psi^{\bar i}_{z \bar z}\\
    \psi_{\bar i}^* & \psi_{\bar i}^* & \psi_{\bar i \bar z}^* & \psi_{\bar i z \bar z}^*
\end{array}
\end{align}

The canonically conjugate variables are written as three pairs of rows, and in each pair
of rows the conjugate fields have an opposite horizontal position.
(If we enumerate the rows from 0 to 5 and the columns from 0 to 3 in this diagram,
then the fields in positions $(2i,j)$ and $(2i+1,3-j)$ are canonically conjugate pairs
of a field and antifield. As usual, they have opposite statistics).
The statistics of the zero forms $x^{i},x^{\bar i}, \psi^*_{\bar i}$ is even (bosonic).
The statistics of the remaining zero forms $\psi_i,x_{\bar i}^* ,
\psi^{\bar i}$ is odd (fermionic). It alternates with the degree
of the differential form on $\Sigma$ in the each row.

Let us make the first step and choose the physical fields $\Phi$
by a box around their symbols
\begin{align}
  \begin{array}{cccc}
   \boxed{ x^i }    & \boxed {x^i_z}  & \boxed{ x^i_{\bar z}} & x^i_{z \bar z}  \\
    \boxed {\psi_i}  & \psi_{iz} & \psi_{i\bar z} & \psi_{i z \bar z} \\
   \boxed{ x^{\bar i}} & x^{\bar i}_{z} & x^{\bar i}_{\bar z} & x^{\bar i}_{z \bar z}\\
   \boxed{ x_{\bar i}^*} & \boxed{ x_{\bar i z}^*} & \boxed{x_{\bar i \bar z}^*}&x_{\bar i z \bar z}^* \\
   \boxed{ \psi^{\bar i}} & \boxed{\psi^{\bar i}_z} & \boxed{ \psi^{\bar i}_{\bar z}} &
\boxed{\psi^{\bar i}_{z \bar z}}\\
    \psi_{\bar i}^* & \psi_{\bar i}^* & \psi_{\bar i \bar z}^* & \psi_{\bar i z \bar z}^*
\end{array}
\end{align}
In each pair of conjugate fields only one field is boxed, therefore the boxed fields
make a Lagrangian submanifold in the full BV phase space.
The boxed fields $\Phi$ are left in the theory after the gauge fixing.
At the second step the unboxed fields
$\Phi^*$ are expressed in terms of $\Phi$ as in~\eqref{eq:LagPsi}
via a suitable gauge fixing function $\Psi(\Phi)$.

Let us take $\Psi(\Phi)$ similar to the ordinary B-model~\cite{Witten:1991zz}
\begin{align}
\Psi = g_{i \bar j} (x^i_z \p_{\bar z} x^{\bar i} + x^{i}_{\bar z} \p_z x^{\bar i}).
\end{align}
That gives
\begin{align}
 x_{\bar i z \bar z}^* = g_{i \bar i}(\p_z x^i_{\bar z} + \p_{\bar z}x^i_z) \\
 \psi^i_{\bar z} = g_{i \bar i}\p_{\bar z} x^{\bar i} \quad  \psi^i_{z} = g_{i \bar i} \p_{z} x^{\bar
 i}.
\end{align}
The remaining unboxed fields are zero on the Lagrangian $L_{\Psi}$. We get the
physical Lagrangian of the B-model\footnote{
The present construction of the B-model differs from~\cite{MR1432574}.
In~\cite{MR1432574} the BV action was written in
terms of $8n$ fields on $\hat \Sigma$ corresponding to the coordinates
in the target space $\Pi T^*\Pi T X$.
The present construction takes the target space $M$ to be
a certain extension of $N = \Pi L$ that has a nondegenerate symplectic
structure. The dimension of the target space $M$ in this work is $6n$ compared to $8n$ in~\cite{MR1432574}.}
\begin{align}
 (\hat \psi_i d \hat x^i +  \hat x_{\bar i}^* \hat \psi^i )|_{L_\Psi} = [g_{i\bar i} ( \p_{\bar z} x^{\bar i} \p_{z} x^{i} + \p_{z} x^{\bar i} \p_{\bar z} x^{i}) +  \psi_i (\p_z x^i_{\bar z} - \p_{\bar z} x^i_{z})
 \psi^{\bar i}] + \\
+ [ g_{i \bar i} \psi^{\bar i} (  \p_z x^i_{\bar z} + \p_{\bar z} x^i_{z} )
+ x_{\bar i}^* \psi^{\bar i}_{z \bar z} + x^*_{\bar i z} \psi^{\bar i}_z+
x^*_{\bar i \bar z} \psi^{\bar i}_z]
\end{align}
The last three quadratic terms can be physically interpreted as auxilary. In the remaining action we
recognize that of~\cite{Witten:1991zz} with a change of notations
$x^i_z \to \rho^i_z, x^i_{\bar z} \to \rho^{i}_{\bar z}, \psi^{\bar i} \to
\eta^{\bar i}, \psi_i \to \theta_i $.
The algebra of physical observables is given by the $Q$-cohomology of
$C^{\infty}(M)$, which is the same as $Q$-cohomology of $C^{\infty}(N)$,
which is the $Q$-cohomology of the algebra of
functions of $(x^i, x^{\bar i}, \psi^{\bar i},
\psi_{i})$.
This is the same space as~\cite{Witten:1991zz} where
the primary observables $\CalO^{(0)}$ were identified with the $Q$-cohomology
of functions of  $(x^{i},x^{\bar i},\eta^{\bar i}, \theta_{i})$.

One can also see that the deformation of the master action of the B-model by
a holomorphic function $f(x)$ corresponds to the deformation
of the operator $Q$ to the topological
Landau-Ginzburg model with the superpotential $f(x)$. The physical observables are identified
with cohomologies of $Q = \bar \psi^i \frac {\p} {\p x^{\bar i}} + \frac {\p f} {\p x^i} \frac {\p} {\p
\psi_i}$. For polynomial $f(x)$ on $X=\BC^n$
the cohomology of $Q$ in degree 0 is the polynomial ring
${\BC[x^i]}$ factorized over the ideal generated by $\partial_i
f(x)$. The ring $\frac {\BC[x^i]} { df(x^i)}$ is the familiar ring
of observables of the topological Landau-Ginzburg model~\cite{Vafa:1990mu}.

\subsection{Observables and deformations}

What are the observables of the model? In the path integral
formulation of a quantum field theory, observables can be
associated with deformations of the action.
If the theory is gauge invariant, then the deformed action must be
again a gauge invariant functional.

Let us see what happens concretely in the BV approach~\cite{MR1432574,Schwarz:1992nx}. After the gauge
fixing, the partition function is the integral over a Lagrangian
submanifold in the BV phase space
\begin{align}
Z = \int_{L} e^{-\frac {1} {\hbar} S[\phi]}.
\end{align}
The gauge invariance of the theory means that $Z$ does not change with a change
of the gauge fixing condition. Thus, the partition function $Z$ has to be invariant under
deformations of the Lagrangian submanifold $L$ in the space of
fields. For any function $f$, the integral $\int_L f$ is invariant
 under such deformations if $\Delta f = 0$. Taking $f = e^{- \frac
 1 \hbar S}$, one obtains BV \emph{quantum master equation} $\Delta e^{-\frac 1 \hbar S}  =
 0$, or
\begin{align}
\label{eq:BVq}
  -\hbar \Delta S + \frac 1 2 \{S,S\}  = 0.
\end{align}
Let us assume that $S_0$ satisfy~\eqref{eq:BVq} and let us consider a deformation
$S_0 \to S_0 + \delta S$. We obtain
\begin{align}
   -\hbar \Delta (S_0+\delta S) + \frac 1 2 \{ S_0 + \delta S, S_0 + \delta S\} = 0
\end{align}
so
\begin{align}
  -\hbar \Delta \delta S + Q \delta S + \frac 1 2 \{\delta S, \delta S\} = 0,
\end{align}
where we used the definition of the operator $Q\cdot  = \{S_0, \cdot\}$.
In the limit $\hbar =0$, the BV quantum master equation
becomes~\emph{the BV classical master equation}
\begin{align}
\label{eq:BVMC}
 Q \delta S + \frac 1 2 \{\delta S, \delta S\} =0.
\end{align}
also well known under the name Maurer-Cartan equation in deformation theory.
Deformation of $S$ corresponds to the deformation of $Q$ according to the definition $Q = \{S, \cdot \}$.
The classical BV equation is the homological property $Q^2=0$.
Not all deformations of $S$ lead to a new theory. Consider deformation of $S$
by a diffeomorphism that preserves the symplectic structure, that
is, by a Hamiltonian vector field. The vector field, generated by a function
$\delta S$, acts on $S$ as $S \to S + \{\delta S, S\}$. We see, that
such diffeomorphisms corresponds to deformations $S$ of the form $Q
\delta S$. The theory is invariant under \emph{Q-exact}
deformations $S \to S  + Q \delta S$. From~\eqref{eq:BVMC}, infinitesimal
deformations have to be \emph{$Q$-closed} $Q \delta S = 0$.
Therefore the physical space of non-equivalent gauge invariant
infinitesimal deformations (=infinitesimal observables) is the cohomology group of the operator $Q$.
On the moduli space of physical non-equivalent theories, infinitesimal
observables can be viewed as vector fields.

\underline{Example.} Consider the point in the moduli space of generalized
complex structures corresponding to the ordinary complex structure -- the B-model.
The space of functions $C^{\infty}(\Pi L)$ is the
space of $\Omega^{-p,q}$ forms $\mu_{\bar i_1 \dots \bar i_q}^{j_1 \dots j_p}$, the operator $Q$ is $\bar \p$, the Laplacian
$\Delta$ is the holomorphic divergence $\p$,
the bracket $\{,\}$ is the Lie bracket on the holomorphic polyvector fields.
The deformations $S$ are functions on $C^{\infty}(\Pi L) =
\Gamma(\Lambda^{p}TX^{10} \otimes \Lambda^{q}T^*X^{01})\equiv \Omega^{-p,q}(X) $.

The case $(p,q) = (-1,1)$ corresponds to an ordinary deformation of the complex structure
by a Beltrami differential $\bar \p \to \bar \p + \mu \cdot \p$. Equation~\eqref{eq:BVMC} is
the requirement that the new Dolbeault operator $\bar \p$ squares to zero.
It is called the Kodaira-Spencer equation.

The other deformations of degree 2, the types $(-2,0)$ and $(0,2)$,
are geometrical deformations of the underlying Lie bialgebroid structure $(L,L^*)$.
Indeed, functions $S$ on $\Pi L$ of degree 2 by the relation
$Q = \{\mu, \cdot\}$ correspond to operators $Q$ of degree 1.
Such an operator defines a Lie algebroid structure on the vector bundle
$L$. Deformations $S$ of other degrees are not described by
a Lie algebroid structure and do not correspond to a generalized complex structure.
The moduli space of generalized complex structures is locally
generated by deformations (observables) of degree two.
The moduli space of deformations of arbitrary degrees is
the \emph{extended} moduli space~\cite{Witten:1991zz,MR1609624}.

Consider a $(-2,0)$ deformation by a bivector field $\beta^{ij}$.
Contrary to the $(-1,1)$ case, the equation
$ \bar \p \beta + \frac 1 2 \{\beta, \beta\} =0$
is now equivalent to two separate equations $\bar \p \beta = 0$
and $\{\beta, \beta\} =0$, since the first term has the grade
$(-2,1)$, while the second term has the grade $(-3,0)$. The
equation tells us that $\beta^{ij}$ is the holomorphic bivector. The
second condition tells us that $\beta^{ij}$ is holomorphic Poisson.
Therefore, the space of $(-2,0)$ deformations is the space
of holomorphic Poisson bivector fields. In section~\ref{StarProduct} we
consider the open B-model deformed by $\beta^{ij}$.

The space of $(0,2)$ deformations is the space of $\bar\p$-closed
$(0,2)$ forms $b_{\bar i \bar j}$.

\subsection{The extended moduli space} \label{the_extended_moduli_space}
Here we will review the geometrical structure of the extended moduli
space.

In~\cite{Witten:1991zz} Witten introduced the notion of the \emph{extended moduli
space}. In~\cite{MR1609624} Barannikov and Kontsevich
showed that the extended moduli space $\CalM$ has a structure of \emph{Frobenius
manifold}: there is a potential $\Phi$ and metric $g_{ij}$ on $\CalM$,
such that the structure functions $C^{i}_{jk} = g^{il}C_{ljk}$, obtained
from the third derivative $C_{ijk} = \p_{ijk} \Phi$, define
an associative product on the space of vector fields.
In~\cite{Li:2005tz} the same property was shown for the moduli space
of generalized complex structure. In~\cite{MR1919435} Manin showed
that the Frobenius structure naturally
appears on the moduli space of deformations of any differential BV algebra\footnote{with
an additional technical requirement similar to the $\bar \p
\p$-lemma}. Since any generalized CY manifold is associated with a
differential BV algebra, the moduli space of generalized complex
structures is also Frobenius manifold. Physically, the structure
function $C_{ijk}$ is the three-point function of the topological
$\CalJ$-model in genus zero.

We consider a quantum theory in the BV formalism $ \int_{L} e^{-\frac 1 h S}$.
The space of BV functionals is differential BV algebra $(\CalA,Q,\Delta,\tr)$,
where $Q$ and $\Delta$ are the corresponding BV operators, defined above. The trace
map is defined in terms of the path integral $\int_{L}$
over some Lagrangian submanifold $L$ in the BV phase space.

Consider a deformation $S_0 \to S_0 + \delta S$. We observed above that
$\delta S$ must satisfy the quantum master equation
\begin{align}
  -\hbar \Delta \delta S + Q \delta S + \frac 1 2 \{\delta S, \delta S\} = 0.
\end{align}
Looking at this equation, one can keep in mind the familiar case of the ordinary B-model,
where $Q = \bar \p$ and $\Delta = \p$. All arguments are parallel.

Let us require that $\delta S$ solves
both the quantum and the classical BV equations, so $\Delta \delta S
=0$.\footnote{
This restriction appeared in~\cite{Bershadsky:1993cx} in
the target-space formulation of the B-model (Kodaira-Spencer
theory) from the string field theory point of view.
The target space fields $\mu\in \Gamma(\Omega^{-p,q})$ in~\cite{Bershadsky:1993cx}
were restricted by the condition $\p A =0$, which was
interpreted as a condition for a string field to be in $b_0$
cohomology.}
Note, that since
$\Delta$ is of degree $-1$, $Q$ is of degree $1$, and $\{,\}$ is of degree $-1$,
then for geometrical deformations $\de S$ of degree $2$ the equations
$ Q \delta S + \frac 1 2 \{ \delta S, \delta S\} = 0 $ and $
 \Delta \delta S = 0$
follow automatically from the BV quantum master equation.
Let us denote $\delta S = \mu$, and $Q = \bar \p$, and $\Delta=
\p$. We still remember that $(\CalA, \bar \p, \p, \tr)$ is an arbitrary
differential BV algebra corresponding to an arbitrary generalized complex structure.
The equations on $\mu \in \CalA$ are written as
\begin{align}
\label{eq:mu_cond}
 & \bar \p \mu + \frac 1 2 \{ \mu, \mu\} =0 \\
 & \p \mu = 0
\end{align}

Infinitesimal deformations $\mu$ are represented by
cohomology classes of $\bar \p$ restricted to the kernel of $\p$.
What about finite deformations? Let $\mu = x + a$, where $x$ is a
harmonic representative of a $\bar \p$-cohomology, and $a$ is a
$\p$-exact correction, that we have to find. The equation on $a$ is
\begin{align}
\bar \p  a + \frac 1 2 \{ x+a , x+a  \} = 0.
\end{align}
Let us use the definition of the BV bracket~\eqref{eq:BV_bracket} in terms of the BV Laplacian
$\p$. Since $x$ is harmonic and $a$ is $\p$-exact, one has  $\p( x + a) =
0$, therefore
\begin{align}
\bar \p a + \frac 1 2 \p( (x+a)\cdot(x+a)) = 0.
\end{align}
The we can get a formula to recursively solve for $a$ order in order in $x$
\begin{align}
\label{rec_equation}
 a = - \frac 1 2 \bar \p^{-1} \p ( (x+a)\cdot(x+a))
\end{align}
This Kodaira-Spencer method was used in~\cite{Bershadsky:1993cx} in the context of  deformations of the
B-model. It it works exactly in the same way for any
differential BV algebra~\cite{MR1919435}.
The harmonic representatives $x$ of the cohomology classes
physically correspond to the external background. The recursive solution
of~\eqref{rec_equation} can be drawn in terms of tree level Feynman
diagrams~\cite{Bershadsky:1993cx}.

The solution $\mu(x)$ exists if we assume that the analogue of $\p \bar \p$-lemma
holds. The solution defines a map from the $\bar \p$-cohomology $H_{\bar \p}^\bullet(\CalA)$ to
elements of the BV algebra which satisfy the classical and the quantum BV master equation.
The master action
$S_0 + \mu$ solves simultaneously the classical and quantum BV
master equations, and thus corresponds to a topological model.

In other words, the solution $\mu(x)$ provides us with linear
coordinates $H_{\bar \p}^\bullet(\CalA)\simeq TM_{S_0}$ on the extended moduli
space. The definition of this linear structure depends on the base point $S_0 \in \CalM$.
The linear coordinates on the classical geometrical moduli
are called \emph{canonical coordinates} in the context of the B-model in~\cite{Bershadsky:1993cx}.

\underline{Example.} Let us consider $X=T^2$.
A generalized complex structure on $T^2$ could be only of the ordinary or symplectic
type. As a reference point, let us take the complex structure with a period $\tau$.
The linear coordinates in $\CalM$ are bosonic for $H^{0,0}$ and $H^{-1,1}$ and fermionic
for $H^{-1,0}$ and $H^{0,1}$.
The space $H^{-1,1}(X)$ describes ordinary deformations of complex structure by Beltrami differentials.
The space $H^{0,0}$ describes constant terms that could be added to the BV master
action. Such constant terms correspond to a dilatation of the BV measure.

A deformation of a complex structure of a CY manifold
$\p_{\bar j} \to \p_{\bar j} + \mu^i_{\bar j} \p_i$ corresponds
to a deformation of the holomoprhic $(n,0)$-form $ \Omega \to e^{-\mu}\Omega$.\footnote{A
generalized complex structure $\CalJ$ defines $+i$-eigenbundle $L = TX^{01} \oplus T^*X^{10}$.
The term $TX^{01}$ corresponds to the $\bar z$ direction in the
tangent space. It is described by the vector field $\p_{\bar i}$. The deformation by $\mu_{\bar j}^i$
is a deformation of the $\bar z$ direction in the tangent bundle. A new $\bar z$
direction $\tilde \p_{\bar i}$
in the old  basis is represented by the vector field $\p_{\bar j} + \mu^{i}_{\bar j} \p_i$.
The new holomorphic one-forms $dz$ must
annihilate antiholomorphic vector fields $\p_{\bar j} + \mu^{i}_{\bar j} \p_i$.
Therefore they can be locally written in the old basis as $dz^{i}
- \mu^{i}_{\bar j} dz^{\bar j}$. Therefore a new holomorphic
$(n,0)$ form $\Omega$ is $e^{-\mu} \Omega_0$ up to a scale.}
%Now $\mu$ includes all degrees, however in this simplest example only deformations of
%the complex structure $\mu^{-1,1} \in H^{-1,1}(X)$ and rescaling $\mu^{0,0} \in H^{0,0}$
%of the canonical pure  spinor   are present.
For $T^2$ let $(x,y)$ be real coordinates on the standard  unit square. The complex structure $\tau_0$
corresponds to the complex coordinate $z = x + \tau_0 y$ and the holomorphic $(1,0)$ form
$\Omega_0 = dz = dx + \tau_0 dy$. The periods are $a_0 = \int_{A} dz = 1$ and $b_0 = \int_{B} dz \tau_0$.
Let $\al \equiv \mu^{0,0}$ and $\be \equiv \mu^{-1,1}$.
After the deformation $\Omega = e^{-\mu} \Omega_0 = e^{-\al}(dz - \beta d\bar z)$.
The new periods are $a = e^{-\al} (1-\beta)$ and $b = e^{-\al} (\tau_0 - \beta \bar \tau_0)$.
The new complex structure is $\tau = \frac {b} {a}  = \frac {\tau_0 - \beta \bar \tau_0} {1-\beta}$.
If we assume that the definition domain of $\tau$ is the upper half plane (neglecting discrete
$SL(2,\BZ)$ transformations), then the domain for $\beta = \frac {\tau - \tau_0} {\tau - \bar \tau_0}$
is the unit disk $D: |\beta| < 1$. The boundary $|\beta| = 1$ is the degenerate complex
structure $\Im \tau = 0$. The scale is parameterized by a complex plane $\BC$.
 The total bosonic moduli space is $D \times \BC$.

In this example, deformations of the complex structure
were mapped to deformations of the canonical holomorphic form $\Omega_0  \to  e^{-\mu}
\Omega_0$. The same can be done in the extended moduli space.
The coordinates are the \emph{periods} $X^{i}(\mu) = \int_{A^i} \Omega$, where $A^{i}$ is a basis of
cycles $H_{\bullet}(X)$. The formula
$X^i (x|\Omega_0) = \int_{A^i} e^{-\mu(x)} \Omega_0$ gives
a \emph{period map} $H^\bullet(L) \to H^{\bullet}(X,\BC)$. The
dimension of $H^\bullet(L)$ and $H^\bullet(X,\BC)$ is the same
(recall that multiplication map by the canonical pure spinor $\Omega_0$
in the generalized CY case provides isomorphism between
$\Lambda^\bullet(L^*)$ and $\Omega^\bullet(X,\BC)$). The formula
$\Omega_0 \to e^{-\mu} \Omega_0$ can be interpreted
as the action in the spinor representation of the group element
$e^{-\mu}$ that corresponds to the algebra element $\mu$.
The bosonic part of $\CalM$ is mapped to $H^{odd/even}(X)$
if  $\Omega_0$ is of the odd/even type.

The dimension counting in the $T^2$ example is as follows. There are two complex bosonic
directions in $\CalM$: deformations $\beta$ of the complex structure and
rescaling $\alpha$ of the canonical pure spinor $\Omega_0$.
They are mapped into $H^1(X)$ and parameterized by two periods $a$ and $b$.

Usually the B-model is considered in $\dim_{\BC} X = 3$. What is special
in the geometry of the extended moduli space if $\dim_{\BC} X = 3$? Let the reference
point be an ordinary CY manifold $X$ with a  holomorphic $(3,0)$-form $\Omega_0$.
The geometrical deformations of the generalized complex
structure (which are of degree 2) and dilatations (which are of degree 0)
are parameterized by $H^{-2,0} \oplus H^{-1,1} \oplus H^{0,2}$ and $H^{0,0}$.
Using the map by $\Omega_0$, this space is isomorphic to
$H^{1,0}\oplus H^{2,1} \oplus H^{3,2}$ and $H^{3,0}$.
If $\dim_{\BC} X=3$, the space of geometrical deformations
of generalized CY structure is the half of all bosonic deformations $H^{odd}(X)$.
Therefore, precisely  in dimension 3, the \emph{geometrical} moduli space of generalized CY
structures can be parameterized by half of all odd periods $X^i =
\int_{A_{i}}{\Omega}$, where $A_i$ is a basis in $H_{odd}(X)$.

In the case of \emph{ordinary} CY structures in dimension 3, this is a very well-known parametrization
by \emph{half of all 3-cycles} $A_i$ that make basis in $H_{3}(X)$
\begin{align}
  X^i = \int_{A^{i}} \Omega_0.
\end{align}
The complementary periods $F_{i}$ are defined for the
dual\footnote{We use the basis $A_i \circ A_j = B_i \circ B_j =0$, $A_i \circ B^j = \delta_i^j$.}
3-cycles $B^i$
\begin{align}
 F_{i} = \int_{B^{i}} \Omega_0.
\end{align}
The $F_i$ are  functions of $X^{i}$.

In the case of \emph{generalized} CY structures in dimension 3, one needs to
consider the~\emph{half of all odd cycles}.
We can choose a basis $A_i \circ B^j = \de^j_i$, $A_i\circ A_j = B^i \circ B_j = 0$, where
$A_i,B^j \in H_{1} \oplus H_{3}\oplus H_{5}$ and write
\begin{align}
  X^i = \int_{A^{i}} \Omega \quad F_{i} = \int_{B^{i}} \Omega,
\end{align}
where $\Omega_0$ is the canonical pure spinor on $X$\cite{GCS,Hitchin:2}.
The  $X^i$ can be taken as coordinates on the generalized geometrical
moduli space of $\CalJ$-model. As usual, one can get rid of scale and
consider the projective coordinates $(X^0:X^1:\dots:X^{h_{1,0}+h_{2,1}+h_{3,2}})$.
Alternatively, instead of $X^i$, one can use $(\R X^i,\R F_i)$.
It was shown by Hitchin~\cite{Hitchin:2}
that this parametrization of generalized CY structures by $H^{odd}(X,\BR)$ is not degenerate.
He also constructed a certain explicit map\footnote{It is a critical point of a certain
functional. See~\cite{Pestun:2005rp} for a one-loop study of that functional.}
that takes a real 1+3+5-form $\rho$ and
gives a real 1+3+5-form $\hat \rho$, such that the spinor $\Omega = \rho + i \hat \rho$
is pure and closed. Therefore $\Omega$ defines a generalized complex structure.

Moreover, Hitchin showed~\cite{Hitchin:2} that for $\dim_{\BC} X =3$
 the moduli space of~\emph{generalized CY structures} is a special Kahler manifold.
A special Kahler geometry~\cite{MR1695113,MR1972422} is a familiar attribute of $\CalN=2$
theories~\cite{deWit:1984pk,Strominger:1990pd} and
integrability~\cite{Donagi:1995cf,Gerasimov:2004yx}.

Intrinsically,
a \emph{special Kahler}~\cite{MR1894078,MR1972422,MR1695113,Strominger:1990pd}
structure\footnote{A special Kahler geometry
is sometimes called ``rigid special'' Kahler geometry (global $\CalN=2$), while a projective special
Kahler geometry is called ``local special'' Kahler geometry (local $\CalN=2$). A projective
special Kahler can be obtained from a special Kahler that admit a certain $C^{*}$-action.}
on a Kahler manifold $(M,J,\omega)$ is defined as a flat torsion-free connection $\nabla$ on
$TM$ such that (i) the symplectic structure is covariantly constant  $\nabla \omega = 0$,
and (ii) the complex structure
 is $d_{\nabla}$-closed $d_{\nabla} J = 0$. This definition
is related with the commonly used extrinsic
definition of a special Kahler geometry by means of a holomorphic prepotential $\CalF$ as
 follows. Consider a $2n$-complex dimensional
space $V=\BC^{2n}$ with a canonical holomorphic symplectic form
\begin{align}
\Omega = d X^{i} \wedge dF_{i}
\end{align}
where $(X^i,F_i)$ are coordinates on $\BC^{2n}$.
Consider a Hermitian form $h(u,v) = -i \Omega(u, \bar v)$ for $u,v \in V$.
Then any special Kahler manifold $M$ locally is realized as
a submanifold $\phi: M \to \BC^{2n}$ such
that (i) it is holomorphic, (ii) the restriction $\phi^*h$ of the Hermitian form  on $M$ is nondegenerate,
and (iii) it is holomorphic Lagrangian $\phi^* \Omega = 0$.
The condition (i,ii) gives a Kahler structure $\phi^*h$ on $M$, the condition (iii)
gives the \emph{special} property of the Kahler structure.
In real coordinates $(x^i,y^i,u^i,v^i)$, where $X^i = x^i+iy^i$ and $F_i = u_i + i v_i$ we have
the Kahler form $\omega = -\I \phi^* h = dx^i \wedge du_i + dy^i \wedge dv_i$. Since $M$ is Lagrangian,
$\phi^* \Omega = 0$, and since $\R \Omega = dx^i \wedge du_i - dy^i \wedge dv_i$,
the Kahler form on $M$ is
\begin{align}
\omega =2 dx^i \wedge du_i.
\end{align}
Since $d u_i = \R d F_i = \R (F_{ij} dz^j) = \R F_{ij} dx^i - \I F_{ij} dy^j$, we
get
\begin{align}
\omega = -2 \I F_{ij} dx^i \wedge dy^j = - i \I F_{ij} dz^i \wedge \overline{ dz^j}.
\end{align}
The symplectic form $\omega$ on $M$ can be also expressed in terms of the Kahler potential
$\omega = i \p \bar \p K$, where
\begin{align}
K = - \I ( F_i \overline{ X^i}).
\end{align}
The real part of the Kahler form $\R \phi^* h$ is the Kahler metric on $M$.
In real flat coordinates $(x = \R X ,u = \R F)$
the  metric  $g_{ij}$ is  given by the second derivative $g_{ij} = \p_i \p_j K$.

The real coordinates $x^i = \R X^i, u_i = \I F_i$
 define a flat symplectic torsion-free connection $\nabla$ on $M$ by
the condition $\nabla dx^i = 0, \nabla du_i = 0$.
This connection can be  extended  to the complexified tangent bundle $TM \otimes \BC$, and
then $(0,1)$ part of $\nabla$ is the $\bar \p$-operator $\nabla = \bar \p$,
and  Christoffel symbol of $(1,0)$ part of $\nabla$
is $\p_k \tau_{ij} = \p_{kij} \CalF$, so this is the
familiar connection on the moduli space of
an ordinary topological B-model~\cite{Bershadsky:1993cx}.
The moduli space of generalized CY structures for $\dim_{\BC} X = 3$
has exactly the same geometry.

Locally, a holomorphic Lagrangian submanifold $M \subset \BC^{2n}$ is defined
by a holomorphic function $\CalF(X_i)$ and equations $F_{i} = \frac {\p \CalF}{\p X_i}$.
The  $X^{i}$ are holomorphic coordinates on~$M$. They are called
special coordinates~\cite{Bershadsky:1993cx}.

To summarize, in complex dimension 3, the \emph{geometrical moduli space} of
the generalized $\CalJ$-model is precisely half-dimensional subspace of the
bosonic part of the \emph{extended moduli space}. This half-dimensional subspace
is realized as a holomorphic Lagrangian submanifold with respect to
the canonical holomorphic structure in $H^{odd/even}(X,\BC)$.
It is Lagrangian, because in complex dimension 3,
the canonical symplectic form\footnote{$\omega(\mu_1,\mu_2)  = \tr \mu_1 \mu_2$ for $\mu_1,\mu_2
\in\CalA$. It is nonzero only if $\deg \mu_1 + \deg \mu_2 = 2n$, where $\dim_{\BC} X = n$. The
geometrical deformations by definition are of degree 0 and 2.
That is a half of all bosonic deformations of degrees 0,2,4,6 if $\dim_{\BR} X = 6$. And
that is a Lagrangian submanifold since the degrees 0 and 2 cannot make 6, thus the
symplectic structure on $\CalM$ vanishes.} on $\CalM$ vanishes for geometrical deformations.

Note, that the definition of \emph{special}  and  \emph{canonical} coordinates is
very different. The canonical coordinates $x$ are defined
in tangent space to the extended moduli space
 $T\CalM_{\Omega_0}$ at the reference point $\Omega_0$. A finite
deformation is given by solution $\mu(x)$ of the
BV-Maurer-Cartan equation~\eqref{eq:mu_cond}.
As we will see, the holomorphic metric $\tr \mu_a \mu_b$ in the extended moduli space is
a constant matrix in canonical coordinates.

On the other hand, the \emph{special} coordinates are defined only for $\dim_{\BC} X = 3$
and only for \emph{geometrical} deformations of generalized CY structures.
In this case the geometrical moduli space is the half-dimensional
holomorphic Lagrangian submanifold of the extended moduli space. The special coordinates
are convenient to compare the generalized $\CalJ$-model and the $\CalN=2$
four-dimensional supergravity in the type IIA/B string
compactifications on $X$~\cite{Grana:2005ny,Grana:2005sn,Benmachiche:2006df,Grana:2005jc,Grana:2004bg,Pestun:2005bh}.

There are other related reasons why the six real dimensions is a special case.
In~\cite{Nekrasov:2004vv} Nekrasov
conjectures $Z$-theory (a topological analogue of $M$-theory), which should provide
a non-perturbative completion of topological strings. It is possible that
this theory should be formulated in terms of some $G_2$ theory
on a real seven dimensional manifold $X_7$.
The six-dimensional theory on $X_6$ arises
after compactification of one dimension for $X_7=S^1 \times X_6$.
See also~\cite{Dijkgraaf:2004te,Gerasimov:2004yx}.

Geometrically, infinitesimal observables of the $\CalJ$-model
are vector fields $\Gamma(T\CalM)$.
Using the trace map of the BV algebra $\CalA$ we can define $n$-point function of observables.
Let $x^i \in H^\bullet_{\bar \p}(\CalA) \simeq
T\CalM_{\Omega_0}$ be local coordinates on the moduli space
$\CalM$ near the point $\Omega_0 \in \CalM$.
The $n$-point function of vector
fields $v^i_{1},\dots,v^i_{n}$ is defined in terms of the trace map $\tr$ and the
map $\mu(x)$
\begin{align}
\label{n-point}
 \langle v_{1},\dots,v_{n} \rangle= v_{1}^{i_1} \dots v^{i_n}_n
\tr \p_{i_1} \mu(x) \p_{i_2} \mu(x) \dots \p_{i_n} \mu(x).
\end{align}

The one-point function defines the identity vector field, the two point function
defines the metric on $\CalM$, the three-point function defines the Frobenius
structure~\cite{MR1919435}.
In the \emph{canonical} coordinates $x$ the metric is a  constant matrix
\begin{align}
g_{ij} = \tr \p_i \mu(x) \p_j \mu(x) = \tr \p_i (x + a(x)) \p_j (x+a(x))  = \tr \p_i x \p_j x
\end{align}
The terms with $a(x)$ vanish since they contain a product of $\p$-exact and $\p$-closed
terms, but  $\tr (\p a) b =
\pm  \tr a (\p b)$. The three-point function is not generally constant.

The $n$-point function is the $n$-point function of the toy version
of the $\CalJ$-model on the zero-dimensional worldsheet -- point.  The space of maps
is simply the target space $M$, and the BV algebra of the model is the BV algebra $\CalA$ of functions on $M$.

Of course, the $\CalJ$-model on a two-dimensional worldsheet is
more complicated. Instead of $M$ one has to
consider the space $\Maps(\hat \Sigma, M)$.

We begin with study of topologically trivial maps in genus zero.
By topologically trivial we mean maps which are homotopic to constant maps.
The three-point function is given by the same formula~\eqref{n-point},
since the moduli space of Riemann surfaces of genus zero with three marked points
is a point, and the BV path integral is reduced to the integral
over~$M$. Let us consider the simple case in more details.

%the same
%since
%is the BV algebra $\CalA$
%associated with generalized CY structure on.
% $\CalJ$-model
%with gravity on the worldsheet, the structure of correlation functions essentially
%changes. In genus zero only a three-point function does not vanish at the origin
%of canonical coordinates $x=0$,
%since  Therefore, computation of a three-point function  in $\CalJ$-model
%in genus zero can be done without complications arising from coupling
%with two-dimensional topological
%gravity degrees of freedom. Presumably, the formalism of Sen and Zwiebach~\cite{sen-1996-177,sen-1994-423} might
%be used to couple the theory with the two-dimensional gravity.

\subsection{The topologically trivial maps in genus zero}

This problem was essentially solved
already in~\cite{Bershadsky:1993cx} where the Kodaira-Spencer
action, evaluated in a critical point, was suggested as a free energy
of B-model in genus zero. Geometrically, the moduli space of the $\CalJ$-model is a Frobenius
manifold. The construction was generalized for the
the extended moduli space of B-model in~\cite{MR1609624}, and
for the moduli space of any differential BV algebra
in~\cite{MR1919435}. Generalized CY structure is associated with
a differential BV algebra, so by the construction
of~\cite{MR1919435} the moduli space of generalized CY structures
is also Frobenius. Thus, it is automatically equipped with a certain potential
function $\Phi$. The derivative $\Phi_{ijk}$ is the three-point
function, which was defined above in terms of the trace map $\tr$ and
$\mu(x)$-map. In~\cite{Li:2005tz} Li directly studied the
moduli space of generalized CY structures.

Let us review the generalized Kodaira-Spencer construction.
Let $\Omega_0$ in $\CalM$ be the reference point and
$x \in T\CalM_{\Omega_0} =  H^\bullet(L_{\Omega_0})$ be the
canonical coordinates.
The three-point function $C_{ijk}$ is defined as
\begin{align}
\label{eq:3pt_func}
C_{ijk}(x |\Omega_0) = \tr_{\Omega_0} \mu_{i} \mu_{j} \mu_{k},
\end{align}
where $\mu_{i} \equiv \frac {\p \mu(x)} {\p x^i}$. The index $\tr_{\Omega_0}$ is written to remember
that the trace map~\eqref{eq:trace_map} is defined in terms of
the canonical pure spinor $\Omega_0$ of the \emph{reference} point in $\CalM$.

Is there a potential function $\Phi(x|\Omega_0)$ such that the three-point function is the third
derivative? That is
\begin{align}
C_{ijk}(x |\Omega_0) = \p_{i} \p_{j} \p_{k} \Phi(x|\Omega_0).
\end{align}

The answer is  very natural from the BV point of view~\cite{Witten:1990wb}.

Namely, let us consider the classical BV master equation
\begin{align}
\label{eq:classBV2}
 \bar \p \mu + \frac 1 2 \{\mu, \mu\} = 0
\end{align}
for deformations of the master action $S_0 \to S_0 + \mu$.
We can think about this equation in analogy with the flat curvature equation
$ dA + A \wedge A =0 $ for a connection $A$.
What is the functional on a space of  connections
 that has the equation of motion $dA + A \wedge A =0$?
This is the Chern-Simons~\cite{Witten:1992fb,Axelrod:1989xt} functional
\begin{align}
 S[A] = \int \frac 1 2 \tr A d A + \frac 1 6 \tr A [A,A].
\end{align}
For any special differential BV algebra we can consider
Chern-Simons like functional\footnote{The functional on space of BV actions!},
\begin{align}
 S_{CS}[\mu|\Omega_0] = \frac 1 2 \tr  \mu \bar \p  \mu + \frac 1 6 \tr \mu \{\mu,\mu\}
\end{align}
whose equation of motion is the BV master
equation~\cite{Witten:1990wb}.
Let us impose the constraint $\p \mu = 0$~\eqref{eq:mu_cond}, which makes   $\frac{1}{\p}$
to be well defined. Then $\{\mu, \mu\} = \p (\mu \cdot \mu)$. We can pull out
$\p$ and get the action
\begin{align}
\label{eq:BV_CS}
 S_{CS}[\mu|\Omega_0] = \frac 1 2 \tr  \mu \frac {\bar \p}{\p} \mu + \frac 1 6 \tr \mu \mu
 \mu.
\end{align}
The equation of motion of~\eqref{eq:BV_CS} is the BV classical master equation~\eqref{eq:classBV2}.
This action in the context of an ordinary B-model
was written in~\cite{Bershadsky:1993cx} and was called
Kodaira-Spencer target space action. Here, as in~\cite{MR1919435},
it is written in the context of any differential BV algebra $(\CalA,\bar \p, \p, \tr)$.
The case of generalized CY structures is a particular case of the general construction.
The abstract generalized Chern-Simons theory was studied
in~\cite{Schwarz:2005rv}.

Let us rewrite~\eqref{eq:BV_CS} more carefully. Let us assume that
the BV algebra $\CalA$ satisfies the analogue of the $\p\bar \p$-lemma~\cite{cavalcanti-2005-}.
Let $x \in H^\bullet(\CalA)$ be  harmonic, so $\bar \p x = 0, \p x = 0$.
Let us represent the $\p$-closed element $\mu\in \CalA$ as $\mu=x+a$, where $a$ is $\p$-exact.
Using the relation $\{\mu,\mu\} = \p (\mu \cdot \mu)$ for $\p$-closed $\mu$ we get
\begin{align}
\label{eq:BV_CS2}
 S_{CS}[a|x,\Omega_0] = \frac 1 2 \tr a \frac {\bar \p} {\p}  a + \frac 1 6 \tr (x+a)\cdot (x+a) \cdot (x+a).
\end{align}
As explained in the previous subsection,
the solution $\mu(x) =  x + a(x)$ of the equation~\eqref{eq:classBV2}
for a critical point of~\eqref{eq:BV_CS2}
can be found perturbatively. By the standard field theoretical
argument, the solution is a generating function
of tree Feynmann diagrams of the cubic theory~\eqref{eq:BV_CS2}.
The vertex in these diagrams is $\tr \mu \mu \mu$, the propagator is $\bar \p^{-1} \p$.
The three-point function $C_{ijk}$ is the third derivative
of the action~\eqref{eq:BV_CS2} evaluated at the critical point $a(x)$
\begin{align}
\label{eq:BV_CS_3pt}
 C_{ijk}(x|\Omega_0) = \p_i \p_j \p_k \Phi(x|\Omega_0) \\
 \Phi(x|\Omega_0) = S_{CS}[a(x)|x,\Omega_0]
\end{align}
In different, but similar contexts, that was shown in many places
like~\cite{Bershadsky:1993cx,MR1609624,MR1919435}. Let us briefly remind
the computation. Let $\dot f(x) = \frac {d} {d \tau}f(x)$
be the derivative in an arbitrary constant direction $\p_{\tau} = v^{i} \p_i$
 in $H^{\bullet}(\CalA)$.
We need to show that
\begin{align}
 \dddot \Phi(x)  =  \tr \dot \mu^3(x)
\end{align}
We have
\begin{align}
 \dddot \Phi = ( \frac 1 2 \tr a \bar \p \p^{-1} a  + \frac 1 6 \tr \mu \mu \mu)\spdddot
\end{align}
Let us compute derivatives
\begin{align}
\frac 1 6 (\tr \mu^3)\spdddot =  \tr \dot \mu^3 + 3 \tr  \ddot \mu \dot \mu \mu +
\frac 1 2  \tr  \dddot \mu \mu^2 \\
\frac 1 2 (\tr a \bar \p \p^{-1} a)\spdddot =
3 \tr \ddot a \bar \p \p^{-1} \dot a + \tr \dddot a \bar \p \p^{-1} a
\end{align}
Since $\mu = x + a$, we have $\ddot \mu = \ddot a$, and $\dddot \mu = \dddot a$, therefore
\begin{align}
 \dddot \Phi(x) = \tr \dot \mu^3 + 3 \tr \ddot \mu ( \bar \p \p^{-1} \dot a + \mu \dot \mu) +
\tr \dddot \mu (\bar \p \p^{-1} a + \frac 1 2 \mu^2)
\end{align}
The factors in the brackets of $\ddot \mu$ and $\dddot \mu$ terms
vanish at the critical point~\eqref{eq:classBV2}.
Indeed, starting from~\eqref{eq:classBV2},
we have $
\bar \p \p^{-1} a + \frac 1 2 \mu \cdot \mu = 0$,
and taking the derivative $\p_{\tau}$ we get
$\bar \p {\p^{-1}}\dot a + \mu \dot \mu=0.$

Thus, for any special differential BV algebra, in particular, for any
that comes from a generalized CY manifold, there is a secondly quantized
theory of the Chern-Simons type. In more details this theory was studied in~\cite{Losev:0506039}.

There is one interesting remark about loop computations in~\eqref{eq:BV_CS}.
Counting degrees in the cubic vertices and propagators,
one can see that there is the  selection rule for nonzero correlation functions
\begin{align}
\label{eq:CS_degree}
 \frac 1 2 \sum_{i=1}^{n} d_i = (\dim_{\BC} X - 3)(1-g) + n.
\end{align}

(Proof. Let $D = \dim_{\BR} X$, the degree of $\mu_i \in \Lambda^{k}(L^*)$
be $k$.
Let us mark degrees at the incoming line of each vertex. Each propagator
has two marks on its ends. In each vertex we have $ d_{j1}+d_{j2}+d_{j3} = 2D$,
where $j_1,j_2,j_3$ enumerate the incoming lines to that vertex.
In each propagator we have $d_{i1} + d_{i2} + 2 = 2D$, where $d_{i1},d_{i2}$ are degrees
at the ends of a given propagator $i$.  This condition is due to the form $\tr a \p^{-1} \bar \p a$
of the kinetic term, where $\p^{-1} \bar \p$ has degree 2.
Let $d_i$ be degrees of fields in the external lines entering the diagram, $n$ be
the number of external lines, $V$ be the number of vertexes, $E$ be the number
of propagators, $g$ be the number of loops.
 The total sum of degrees over ends of all lines can be computed
as sum over vertexes or the sum over lines. Since the result is the same we have the equation
$2VD = 2E(D-1) + \sum {d_i}$. Plugging relations for a trivalent graph  $V = 2(g-1) + n$ and
$E  = n -3 + 3g$ we get~\eqref{eq:CS_degree}.)

This is precisely the condition on degrees of observables in the A-model
with non-vanishing Gromov-Witten invariants. In Gromov-Witten theory
$ (\dim_{\BC} X - 3)(1-g) + n + \int_{\beta} c_1(TX)$.
is the dimension of the fundamental class in the moduli
space of Riemann surfaces $\CalM_{g,n}$ of genus $g$ and degree
$\beta$.

For maps of degree 0 the Chern-Simons like theory~\eqref{eq:BV_CS} gives the same condition.
Moreover, that condition holds in arbitrary $\dim_{\BC} X$
and for an arbitrary generalized CY structure. It is not specific to the A-model.
That tells us that $\dim_{\BC} X = 3 $ is not crucial for the existence
of the target space theory of the Chern-Simons form~\eqref{eq:BV_CS}. Indeed, we never
used condition $\dim_{\BC} X = 3$ to write the action. See
also~\cite{Schwarz:2005rv}.
The peculiarity of $\dim_{\BC} X = 3$ is that this CS theory is nontrivial in the sector
of purely \emph{geometric} deformations of generalized CY structure.

In the case when $\CalJ$-model is the B-model,
the theory~\eqref{eq:BV_CS} reduces to the Kodaira-Spencer
theory of gravity of~\cite{Bershadsky:1993cx},
in the case when $\CalJ$-model is the $A$-model, it reduces to the
Kahler gravity of~\cite{Bershadsky:1994sr}.

The Chern-Simons theory for BV actions again satisfies the BV master equation~\cite{Bershadsky:1993cx}.
In~\cite{Bershadsky:1993cx} the action was written initially
only for ordinary geometric deformations given by Beltrami differentials,
then by arguments from  string field theory
all other degrees were introduced that had interpretation of BV ghosts
and antifields. Here we obtain the complete Chern-Simons functional with
all degrees included from the very beginning.

In a spirit similar to~\cite{Pestun:2005rp}, one can also
compute the 1-loop partition function in~\eqref{eq:BV_CS2} and find a
certain product  of generalized Ray-Singer torsions in terms of determinants of
 generalized $\bar \p_L, \p_L$ operators  of the Lie algebroid~\cite{Gualtieri:hodge}.

\subsection{The $\dim_{\BC}X = 3$ case }

In case $\dim_{\BC}X = 3$ there exist
additional structure on the moduli space of $\CalJ$-model.
The logic below closely follows the standard considerations of the B-model~\cite{Bershadsky:1993cx}.
The difference is that a holomorphic $(3,0)$ form is replaced by a canonical
pure spinor (a mixed differential form).
The familiar basis of all \emph{3-cycles} is replaced by the basis of all \emph{odd(even)} cycles.

It was explained in section~\ref{the_extended_moduli_space}
that the moduli space of geometrical deformations
is a holomorphic Lagrangian half-dimensional submanifold $L$
in the bosonic slice of the total extended moduli space.
We can parameterize the bosonic extended moduli space $H^{odd/even}(X)$ by
periods of the canonical pure spinor
\begin{align}
  X^I = \int_{A_I} \Omega, \quad F_I = \int_{B^I} \Omega
\end{align}
where $A_I$ and $B^I$ satisfying
$A_I \circ A_I = B^I \circ B^I = 0, A_I \circ B^I = \delta_I^J$ is a basis
of cycles in $H_{odd/even}(X)$. The number of $X^i$ is the half of $\dim H^{odd/even}$,
so we can consider $X$ to be coordinates on the geometric moduli space.
Let us consider a geometric deformation $\mu \cdot \Omega$
of the canonical pure spinor. By degree counting we see that
\begin{align}
  \int_X (\mu \Omega, \Omega) = 0,
\end{align}
so~\footnote{Actually we adjust signs of cycles such that bilinear pairing defined on pure spinors
gives $\int (\Omega_1, \Omega_2) =
\int_{A_I} \Omega_1 \int_{B^I} \Omega_2 - \int_{B_I} \Omega_2 \int_{A^I} \Omega_1$
for two closed pure spinors $\Omega_1, \Omega_2$.} $\dot X^I F_I - X^I \dot F_I = 0$
for any geometrical deformation $\dot X$. A variation along $X^I$ gives
\begin{align}
\label{eq:F_I}
F_I = X^J \p_{I} F_J.
\end{align}
Let us consider the function
\begin{align}
\label{eq:CalF1}
  \CalF(X) = \frac 1 2 X^I F_I
\end{align}
and compute $\p_J \CalF$
\begin{align}
  \p_J \CalF(X) = \frac 1 2 (\p_J F_I) X^I + \frac 1 2 F_J.
\end{align}
Using the equation~\eqref{eq:F_I}, we see that
\begin{align}
  \p_J \CalF(X) = F_J
\end{align}
so $\CalF(X)$ is the generating function of the Lagrangian submanifold $L$
embedded into $H^{odd/even}$ with symplectic coordinates $(X^I, F_I)$.

We see that the geometrical moduli space of the generalized $\CalJ$-model for $\dim_{\BC} X = 3 $
has the familiar special Kahler structure. It is equipped by the
holomorphic potential $F(X)$ and the special coordinates $X^I$.

Let us also note that the third derivative $C_{IJK} = \p_I \p_J \p_K \CalF$ defines
the three-point function that agrees with~\eqref{eq:3pt_func}.
Indeed, consider the third derivative $\dddot \CalF(X)$ along some constant
vector field in  $X^I$ coordinates.
Using the relation $\CalF = \frac 1 2 X^I F_I$ and $F_I = \p_I \CalF$, we get
\begin{align}
 & \dot \CalF = \dot F_I X^I = F_I \dot X^I \\
 & \ddot \CalF = \ddot F_I X^I + \dot F_I \dot X^I = \dot F_I \dot X^I, \quad \ddot F_I X^I = 0,
&\dddot F_I X^I + \ddot F \dot X = 0 \\
&  \dddot \CalF = \dddot F_I X^I + 2 \ddot F_I \dot X^I  = - \dddot F_I X^I
\end{align}
From the last line we obtain
\begin{align}
  \dddot \CalF = \int (\dddot \Omega, \Omega).
\end{align}
To make the relation with the three-point function,
defined in terms of the
$\mu(x)$-map in the canonical coordinates,
let us represent $\Omega = e^{-\mu}\Omega_0$. Then let us compute $\dddot \Omega$
\begin{align}
 (e^{-\mu}) \spdddot  = (-\dddot \mu + 3 \ddot \mu \ddot \mu - \dot \mu^3)e^{-\mu}
\end{align}
Therefore,
\begin{align}
   \dddot \CalF = -\tr_{\Omega} (\dot \mu^3+ 3\ddot \mu \ddot \mu - \dot \mu^3) = -\tr_{\Omega}(\dot \mu^3),
\end{align}
since the second and the third term in the bracket vanish for geometrical deformations by a simple
degree counting.
Explicitly, we can write
\begin{align}
\label{eq:3pt_calF}
  C_{IJK} =\p_I \p_J \p_K \CalF = - \int ( \mu_I \mu_J \mu_K \cdot \Omega, \Omega)
\end{align}
where $\mu_I$ represents the basis for deformations of $\Omega$, that is
$\delta X^I \delta_I \Omega = \mu_I \Omega $. Up to a sign redefinition we
 nearly recovered the standard formula
for~\eqref{eq:3pt_func}
\begin{align}
\label{eq:3pt_Phi}
 \p_i \p_j \p_k \Phi =
 C_{ijk} = \tr_{\Omega^0} \mu_i \mu_j \mu_k = \int (\mu_i \mu_j \mu_k \Omega_0, \Omega_0).
\end{align}
The apparent difference is that~\eqref{eq:3pt_calF} uses the measure $\Omega$
for the trace map
of that point $x$ where $C_{IJK}$ is computed, but~\eqref{eq:3pt_Phi}
uses the measure $\Omega_0$ of some reference point.
However, if $\Omega = e^{-\mu(x)} \Omega_0$ where $\mu(x)$ is variation
of the generalized complex structure by means of $\mu \in
\Lambda^{2}L^*$, without a change of the scale of the canonical pure spinor, then
\begin{align}
\tr_{\Omega^0} \mu_i \mu_j \mu_k = \tr_{\Omega} \mu_i \mu_j \mu_k.
\end{align}
This easy to see using  $\dim_{\BC} X = 3$ and counting degrees.

So we showed, that the formalism of the previous section, with the
Chern-Simons type formulas~\eqref{eq:BV_CS_3pt}~\eqref{eq:BV_CS}
for the partition function of BV actions,
which is valid in any dimension and for all types of deformations,
reduces for $\dim_{\BC}X=3$ and geometrical
deformations $\mu(x) \in \Lambda^{2}L^*$ to the familiar
formulas of the special Kahler geometry~\eqref{eq:CalF1}~\eqref{eq:3pt_Phi}.

It is known that the topological string on a CY threefold $X$ computes the effective
prepotential for vector multiplets of the $\CalN=2$ four-dimensional
supergravity arising after compactification of type $II$ string
theory on $X$~\cite{Bershadsky:1993cx,Antoniadis:1993ze}.

It is also possible to consider compactification of type II string
theory on a generalized CY
threefold~\cite{Witt:2005oct,Grana:2004bg,Grana:2005sn,Grana:2005jc,Grana:2005ny}.
One can expect that the relation between the physical
and topological string also holds for such compactifications.
A generalization of the OSV conjecture~\cite{Ooguri:2004zv} on the relation between black hole entropy in the type II string compactifications and the topological strings is considered in~\cite{Pestun:2005bh}.

We did not discuss the question of the holomorphic anomaly~\cite{Bershadsky:1993cx,Witten:1993ed}
in the generalized complex case. However, as it was demonstrated by A.~Gerasimov and S.~Shatashvili
in~\cite{Gerasimov:2004yx}, the holomorphic anomaly equation
can be defined on an arbitrary moduli space that has a structure of
a special Kahler manifold $M$.
Since the geometric moduli space of generalized complex structures for $\dim_{\BC} X = 3$
is  special Kahler, the holomorphic anomaly  is defined in the same way.
Presumably, the specifics of $\dim_{\BC} X = 3$ and special Kahler geometry is
not important for the holomorphic anomaly equation. The key point of the holomorphic
anomaly equation is a parallel transport of observables on the moduli space.
Given a flat connection on the extended moduli space for generalized $\CalJ$-model,
one should be able to get again the holomorphic anomaly equation.
It would be interesting to see its simple derivation from the target space perspective of
the action~\eqref{eq:BV_CS}.

\section{The Kontsevich $*$-product as a $(-2,0)$ deformation
of the open topological B-model by a holomorphic Poisson bivector\label{StarProduct}}

In this section we consider an example of the topological $\CalJ$-model, which
is not equivalent to the ordinary  A-model or B-model.
More specifically, we will consider $\CalJ$ to be a finite deformation
of the ordinary B-model
by a holomorphic Poisson bivector $\beta^{ij}$.
We will take the B-model as a reference point, and consider perturbation theory over $\beta^{ij}$.
We expect to get the Kontsevich $*$-product formula~\cite{MR2062626,MR1855264}
as a deformation
 of the algebra of the open  B-model in accordance with~\cite{Kapustin:2003sg,Hofman:2000ce,Losev:1997tp}.

The $\beta$-perturbed B-model has the BV master action
\begin{align}
\label{eq:B_open_BV}
S_{B} = \int_{\hat \Sigma} \mu ( \psi_i dx^i + \psi^{\bar i} x_{\bar i}^* + \beta^{ij}\psi_i\psi_j)
\end{align}
Let us stress that $\beta$ is not supposed to be infinitesimally
small. Now we want to consider the action~\eqref{eq:B_open_BV} on a disk.
The observables are holomorphic functions on the boundary: we
consider the space-filling B-brane.

Let us quantize~\eqref{eq:B_open_BV}. We need to choose a
Lagrangian submanifold. In the conjugate pairs of fields $(x^{\bar
i}, x_{\bar i}^*)$ and $(\psi^{\bar i}, \psi_{\bar i}^*)$ we set
to zero all components of $x^{\bar i}$ and $\psi_{\bar i^*}$:
\begin{align}
  \begin{array}{cccc}
    x^{\bar i} & x^{\bar i}_{z} & x^{\bar i}_{\bar z} & x^{\bar i}_{z \bar z}\\
   \boxed{ x_{\bar i}^*} & \boxed{ x_{\bar i z}^*} & \boxed{x_{\bar i \bar z}^*}&
\boxed{x_{\bar i z \bar z}^*} \\
   \boxed{ \psi^{\bar i}} & \boxed{\psi^{\bar i}_z} & \boxed{ \psi^{\bar i}_{\bar z}} &
\boxed{\psi^{\bar i}_{z \bar z}}\\
    \psi_{\bar i}^* & \psi_{\bar i}^* & \psi_{\bar i \bar z}^* & \psi_{\bar i z \bar z}^*
\end{array}
\end{align}
The action for the boxed fields is a decoupled quadratic free action
corresponding to the second term in~\eqref{eq:B_open_BV}. Since it decouples, we will forget about it.

We are left with the pair of conjugate fields $(x^{i},\psi_i)$ and purely holomorphic BV action
for them
\begin{align}
\label{eq:B_open_BV_holo}
S_{B} = \int_{\hat \Sigma} \mu ( \psi_i dx^i + \beta^{ij}(x) \psi_i\psi_j)
\end{align}
This is consistent, because $\beta^{ij}$ is the holomorphic
bivector, so it depends only on $x^{i}$, and the observables are
holomorphic functions on a boundary, so they also depend only on
$x^{i}$. In the BV holomorphic symplectic structure $x^i$ and $\psi_i$ are conjugate variables.

In the real case precisely the action~\eqref{eq:B_open_BV_holo} was studied
by Cattaneo and Felder~\cite{MR1779159,MR1854134}
in the context of the deformation quantization. In that case context $x^i$ are
real coordinates on a real manifold $X$ and $\psi_i$ are coordinate in the fibers of  $T^*X$.
There is the canonical symplectic form for $(x^i, \psi_i)$.
In~\cite{MR1779159,MR1854134} it is shown  that the 2D perturbative diagrams
for the 3-point boundary correlation function $\la h(x(\infty))f(x(0))g(x(1))\ra$
precisely reproduce the terms in the Kontsevich $*$-product formula~\cite{MR2062626}
\begin{align}
 f(x) \star g(x) = f(x) g(x) +  \frac {i \hbar } {2} \p_i f(x) \p_j g(x) \beta^{ij}(x) + \dots.
\end{align}

The complete formula~\cite{MR2062626,MR1779159,MR1854134}
\begin{align}
 f \star g  = fg + \sum_{i=1}^{\infty} \left ( \frac {i \hbar} {2} \right)^n \sum_{\Gamma}
w_{\Gamma} D_{\Gamma} (f \otimes g)
\end{align}
is the perturbative series in $\hbar$. Each term is given by a certain bi-differential operator $D_{\Gamma}$
acting on $f$ and $g$ and associated with a certain Feynman diagram. The weight $w_{\Gamma}$ is the amplitude
of the diagram $\Gamma$.
The derivation, following Cattaneo and Felder~\cite{MR1779159,MR1854134}, goes as follows.

The sum in the formula runs over diagrams $\Gamma$. A diagram of order $n$ has $n+2$ vertices.
There are $n$ vertices labelled  $1..n$ and corresponding to an insertion
of $\beta^{ij}(x)$ in the interior of the disk $\Sigma$.
There are also two distinguished vertices
labelled $L$ and $R$ for insertion of $f(x)$ at two points $0$ and $1$ at the boundary.

The vertices are connected by oriented  propagators.
Precisely two propagators start at each vertex and end somewhere else.
To such a diagram corresponds a bidifferential operator $\Delta_\Gamma$ that acts on $f$ and $g$.
This operator is made of $\beta^{ij}$ and its derivatives $\p_{i1} \p_{i2} \dots \beta^{j_1 j_2}$.
The arrows indicate contraction of indices of $\beta^{ij}$ and derivatives $\p_i$.
The beginning of an arrow stands for an upper index $i$ of $\beta^{ij}$, and the end
of an arrow points to another $\beta(x)$ or $f(x)$ that should be acted by $\p_i$.
The coefficient $w_{\Gamma}$ is given by a certain integral over
position of $n$ points $z_i$ on the upper half plane (or disk) $\Sigma$.
The integrand is a product of $2n$ functions $\phi(z_i,z_j)$ for each connecting arrow in
the diagram.

That is just a Feynman diagram of
the perturbation theory in $\beta$ for the 2D theory on the disk~\eqref{eq:B_open_BV_holo}.
The free action is $\int \psi_i dx^i$.  The propagator connects $\psi^i$ and $x^i$.
Each insertion of $\beta^{ij}\psi_i \psi_j$ gives  two $\psi$'s.
The $\psi$'s should be contracted with $x$'s, so we expand $\beta^{ij}(x)$ and $f(x)$
in $x$ and contract them with $\psi$'s. Pulling out one $x$ from $\beta^{ij}(x)$ or $f(x)$
 and contracting it with $\psi$ is the same as taking a derivative of $\beta^{ij}(x)$ or $f(x)$.

The coefficient of each diagram is given by a certain integral over
position of $n$ insertion points over $\Sigma$. The propagator
$\phi(z,w)$ is the appropriately gauged fixed operator $d^{-1}$.
It has a structure $\phi(z,w) = \log \frac { (z - w)(z - \bar w) } {
(\bar z - \bar w)(\bar z - w) } $. The integrals over positions of
points $z_1,\dots z_n$ of products of these propagators are
computable~\cite{MR2062626}. They are of a topological nature,
since propagator $d^{-1}$ is represented by a closed differential form.

Exactly the same perturbation theory holds for the~\eqref{eq:B_open_BV_holo}.
Indeed, the fields of~\eqref{eq:B_open_BV_holo}, the BV symplectic structure, the Poisson bivector
and the BV measure in the functional integral are precisely holomorphic analogue of
the Cattaneo and Felder data. The $(-2,0)$ finite perturbation of the B-model
 formally reproduces in all orders
the Kontsevich $*$-product formula in the holomorphic context.

See~\cite{Baulieu:2001fi} for discussion of subtleties related with
quantization of the Poisson sigma-model.

\section{Conclusion \label{Conclusion}}

In this paper we presented the topological sigma-model that
depends only on a generalized complex structure $\CalJ$ on the target space.
We employed the AKSZ formalism~\cite{MR1432574} for construction of the $\CalJ$-model.

The closed sector was studied at the tree level without instanton
corrections, reproducing~\cite{Li:2005tz}.
The relevance of the Chern-Simons like functionals
for~\emph{closed} string field theory was discussed. The observables and
correlation functions are defined in agreement
with Barannikov and Kontsevich~\cite{MR1609624}, and Li~\cite{Li:2005tz}.
The special properties of the case $\dim_{\BC} X = 3$ are studied
from the viewpoint of the general framework.

In the open sector we considered generalized complex structure
represented by an ordinary complex structure and holomorphic Poisson
bivector. The product in the algebra of open strings is
deformed into the noncommutative holomorphic Kontsevich $*$-product
with~\cite{Kapustin:2003sg,Hofman:2000ce,Losev:1997tp}. The computation is
completely parallel to the real case of~\cite{MR1779159,MR1854134}.

There were some related developments recently.
In~\cite{Kapustin:2003sg,Kapustin:2004gv,Kapustin:2005uy} it was
suggested to construct the generalized topological $\CalJ$-model
by means of the generalized Kahler geometry, which is described by a
pair of commuting generalized complex structures $(\CalJ,\CalJ')$.
As Gualtieri showed~\cite{GCS}  the generalized Kahler geometry
$(\CalJ,\CalJ')$ is equivalent to the data $(g,b,I_+,I_-)$ with
certain compatibility conditions, where $g$ is an ordinary metric,
B is a two-form, $I_+$ and $I_-$ are  ordinary complex
structures. (This is the geometry discovered in studies of
$\CalN=(2,2)$ CFT's in~\cite{Gates:1984nk}, see also recent
works~\cite{Zabzine:2005qf,Lindstrom:2004hi,Lindstrom:2004iw,Zabzine:2004dp,Lindstrom:2004eh,
Zucchini:2005rh,Zucchini:2004ta,Zucchini:2005cq,Lindstrom:2005zr,Bergamin:2004sk}.)

The difference of our approach is that from the very beginning we
require existence of only one generalized complex structure
(and a canonical pure spinor, i.e. a weak generalized CY
structure). A second generalized complex structure is used only to gauge fix the model.
In general, it does not have to be integrable. It would be very
interesting to compare explicitly the gauge fixed version of
$\CalJ$-model with constructions
of~\cite{Zucchini:2005cq,Lindstrom:2005zr,Bredthauer:2006hf}.
It is also possible, that one could find other
physically interesting gauge fixing conditions. For example, the
gauge that was used in section~\ref{StarProduct} to study $\beta^{ij}$
deformations of the B-model, is different.
It is not yet clear, what is the space $\CalL$ of
physically non-equivalent gauges -- Lagrangian submanifolds in the BV phase space of
$\CalJ$-model).

The interesting questions about possibility of global symplectic
realization~\cite{MR1747916,MR866024,MR854594,Bojowald:2001ae} of $\Pi L$, as well as about behavior of
the model at the  non-regular points, were left out for the future
study. The topological sigma-model on a generalized complex target
space might be useful in the context of the field theoretical study of geometrical Langlands
program~\cite{Kapustin:2006pk} and generalized type II string
compactifications~\cite{Grana:2005sn}.

In~\cite{Schwarz:2000ct} a generalization of BV formalism was
suggested. There is a way to define amplitudes for cycles in the moduli
space $\CalL$ of Lagrangian submanifolds. The original BV construction
is a particular case of zero cycles. The higher genus amplitudes of the $\CalJ$-model coupled
with the 2D gravity correspond to the integrals of closed forms over cycles in $\CalL$.
One needs to study these higher genus contributions.
There are some indications~\cite{Losev:0506039,Schwarz:2005rv} that the generalized CS
theory of BV actions might define higher genus amplitudes in the no instanton approximation.
The relation~\cite{Bershadsky:1993cx} with the closed string field theory needs to
be studied in more details. In particular, one has to find out, what are the
descendants from the viewpoint of the generalized CS theory.

The generalized CS theory~\eqref{eq:BV_CS} on the space of BV functionals exists for any
theory formulated in the BV way. The classical solutions of this CS
theory correspond to the solutions of the BV classical master
equation. What is the meaning of the generalized CS action for BV functionals at the quantum level?

{\bf Acknowledgements}

I would like to especially thank E.~Witten for coordination of the project,
many interesting discussions, important comments and suggestions.
I am very grateful to N.~Nekrasov for fruitful discussions, ideas
and numerous answers to my questions.
I thank M.~Grana, F.~Denef, A.~Kapustin, A.~Losev, A.~Neitzke,
S.~Shatashvili,  D.~Shih and A.~Tomassielo for interesting discussions and remarks.
Part of this research was done during my visits
to Institut des Hautes Etudes Scientifiques, Bures-sur-Yvette, France and
the 3rd Simons Workshop in Mathematics and Physics at Stony Brook University, NY.
I thank these institutions for their kind hospitality.
The work was supported in part by grant RFBR 04-02-16880 and grant NSF 245-6530.

\bibliography{bsample}

\end{document}